\documentclass{aa}  

\usepackage{graphicx}
\usepackage{txfonts}
\usepackage{amsmath}
\usepackage{xcolor}
\usepackage{subfig}
\usepackage{xspace}
\usepackage{hyperref}
\usepackage{bbold}
\usepackage{lscape}
\usepackage{multirow}

\begin{document} 

   \title{The accuracy of mutual potential approximations in simulations of binary asteroids} 
      
   \author{Alex Ho
          \inst{1}
          \and
          Margrethe Wold\inst{1}
          \and
          Mohammad Poursina\inst{1}
          \and 
          John T. Conway\inst{1}
          }

   \institute{Department of Engineering Sciences, University of Agder, Jon Lilletuns vei 9, N-4879 Grimstad\\
              \email{alex.ho@uia.no}
             }

   \date{Received xx; accepted xx}

\abstract
{Simulations of asteroid binaries commonly use mutual gravitational potentials approximated by series expansions, leading to truncation errors, and also preventing correct computations of force and torque for certain configurations where the bodies have overlapping bounding spheres, such as in the rotational fission model for creating asteroid binaries and pairs.}
{We address errors encountered when potentials truncated at order two and four are used in simulations of binaries, as well as other errors related to configurations with overlapping bounding spheres where the series diverge.} 
{For this we utilize a recently developed method where the gravitational interaction between two triaxial ellipsoids can be calculated without approximations for any configuration. The method utilizes surface integration for both force and torque calculations, and is exact for ellipsoidal shapes. We also compute approximate solutions using potentials truncated at second and fourth order, and compare these with the solutions obtained with the surface integral method. The approximate solutions are generated with the ``General Use Binary Asteroid Simulator'' (\texttt{gubas}).}
{If the secondary is located with its centroid in the equatorial plane of the primary, the error in the force increases as the secondary is moved closer to the primary, but is still relatively small for both second and fourth order potentials. For torque calculations, the errors become more significant, especially if the other body is located close to one of the extended principal axes. On the axes themselves, the second order series approximation fails by 100\%. For dynamical simulations of components separated a few primary radii apart, the fourth order approximation is significantly more accurate than the second order. Furthermore, because of larger errors in the torque calculations, the rotational motion is subject to greater inaccuracies than the translational motion. For configurations resembling contact binaries where the bounding spheres overlap, the errors in both force and torque in the initial stages of the simulation are considerable, regardless of approximation order, because the series diverge. A comparison of the computational efficiency of the force and torque calculations shows that the surface integration method is approximately 82 times and 4 times slower than the second and fourth order potentials respectively, but approximately 16 times faster than the order eight potential. Comparing the computation efficiency of full simulations, including the calculations of the equations of motion, shows that the surface integration scheme is comparable with \texttt{gubas} when an order four potential is used.}
{The errors generated when mutual gravitational potentials are truncated at second or fourth order lead to larger errors in the rotational than in the translational motion. Using a mathematically exact method for computing forces and torques becomes important when the bodies are initially close and the bounding spheres overlap, in which case both the translational and rotational motion of the bodies have large errors associated with them. For simulations with two triaxial ellipsoids, the computational efficiency of the surface integral method is comparable to fourth order approximations with \texttt{gubas}, and superior to eight order or higher.}


   \maketitle

\section{Introduction}
\label{sec:Introduction}

Dynamical simulations of binary asteroids are relevant for understanding their formation processes, and their evolution over time. Among the Near-Earth population of asteroids, smaller asteroids get spun up by the YORP effect, and break apart when they reach a critical spin limit via rotational fission \citep{1980Icar...44..807W, Margot_etal_2002, 2007Icar..190..250P}. Studying the dynamics of post-fissioned asteroids allows us to understand how such systems evolve over time, as they can either remain in stable orbits about each other (as binaries or multiple systems), re-impact, or undergo mutual escape thus forming asteroid pairs \citep{2010Natur.466.1085P, 2011Icar..214..161J, 2016MNRAS.461.3982B, 2020PSJ.....1...25D, 2022alexpaper}. 

Recently, the NASA planetary defense mission Double Asteroid Redirection Test (DART) \citep{2018P&SS..157..104C, 2021PSJ.....2..173R}  was successfully carried out, and simulations of the (65803) Didymos binary system are essential to fully understand and interpret the dynamics before and after the impact, e.g.\ changes on the surface of the secondary, and changes in the orbit of the secondary. 

Asteroids have non-spherical shapes, and for binaries where the components are close, there is a continuous exchange of energy and angular momentum, leading to a significant spin-orbit coupling. The dynamics of non-spherical binary asteroids is  described by the solution to the full two-body problem (F2BP) for rigid bodies \citep{1995CeMDA..63....1M}. However, solving the F2BP is not a trivial task as there is no analytical solution of the mutual potential between two non-spherical bodies (apart from between two thin disks \citep{Conway_2016, 2021CeMDA.133...27W}). Therefore, approximations of the mutual potential are often made in order to solve the F2BP.

The most common approach is to expand the potential through the use of spherical harmonics \citep{SCHEERES199667, 2002JGCD...25..765H, 2016MNRAS.461.3982B, 2017AJ....154...21F}. However, a drawback with this is that the power series only converges outside the bounding sphere (Brillouin sphere), which is the smallest sphere that can completely contain the circumference of the body. Inside the Brillouin sphere, the power series approximating the gravitational potential diverges \citep{Moritz_1980}. Furthermore, when using series approximations to the potential, higher-order terms are often neglected, which leads to 
truncation errors. 

\citet{1988CeMec..44...49P} presents a different approach to compute the mutual potential of two bodies of finite sizes using power series. The mass distribution of the bodies is described through inertia integrals. Unlike spherical harmonics, where the potential is expressed in spherical coordinates, the method by \citet{1988CeMec..44...49P} uses Cartesian coordinates. \citet{2008CeMDA.100..319T} has further applied this method to bodies with arbitrary shapes and mass distributions. The power series of \citet{1988CeMec..44...49P} converts the six-dimensional volume integral of the mutual potential to six open summations. The method was improved by \citet{2017CeMDA.127..369H}, who reduced the six open summations to one single summation in order to make it more computationally efficient. Moreover, the formulation of \citet{2017CeMDA.127..369H} allows the inertia integrals to be stored before the mutual potential is calculated, while the approaches by \citet{1988CeMec..44...49P} and \citet{2008CeMDA.100..319T} require the inertia integrals to be re-computed before the mutual potential is evaluated. Even though this approach is computationally very efficient, it will still suffer from divergence problems within the bounding spheres of the bodies \citep{2008CeMDA.100..319T}, and truncation errors. 

The potential of an arbitrarily shaped body can also be modeled with a homogeneous polyhedron. This approach has been utilized to determine the gravitational potential of an asteroid \citep{1997CG.....23.1071W, Werner_Scheeres_1997, 2012Geop...77F...1T, Conway_2015}, and the mutual potential of two polyhedra \citep{2005CeMDA..91..337W, 2006CeMDA..96..317F, 2008Icar..194..410F, 2006Sci...314.1280S}. However, this method can be very time consuming if the polyhedron is represented by many triangular faces. A different method is the mascon model \citep{1968Sci...161..680M, Muller1969ConsistencyOL, Geissler_etal_1996}, where the body is modeled by point masses to represent its mass distribution. On the other hand, despite using many point masses, the mascon model can yield large errors in the forces and the resolution of the surface is poorly represented by spherical balls \citep{Werner_Scheeres_1997}. The mascon model has been modified to become more accurate by replacing the point masses with tetrahedrons \citep{2015MNRAS.450.3742C, 2017MNRAS.464.3552A, 2020MNRAS.496.1645A, 2021RoAJ...31..241A}, which provides the gravitational potential to that of a polyhedron. \citet{2015MNRAS.450.3742C} found that their method results in more accurate estimation of the gravitational potential close to the body, and is also computationally faster, compared to the polyhedron approach by \citet{2012Geop...77F...1T}. \citet{2021RoAJ...31..241A} showed that the approach by \citet{2015MNRAS.450.3742C} reduced the computation time by more than 95\%, while losing less than 2\% of the precision, compared to the polyhedron approach outlined by  \citet{2001Geop...66..535T}.

Previous works that have studied the dynamics of asteroid binaries after fission typically consider a mutual potential approximation order of order two \citep{2009CeMDA.104..103S, 2011Icar..214..161J, 2016MNRAS.461.3982B}, or order four \citep{2020PSJ.....1...25D}. However, only a few authors have considered the significance of higher order terms in simulations, and how the order of the series approximation affects the dynamics. \citet{2017CeMDA.127..369H} investigate the importance of higher order terms for a planar two-ellipsoid system where the ellipsoids are initially in contact. They find that truncating the potential at second order is sufficient to describe systems where the mutual orbit is Hill stable, and also when the bodies undergo mutual escape. On the other hand, they find that additional terms become necessary to describe the trajectory if the bodies are highly elongated. \cite{2020PSJ.....1...25D} find that higher-order terms in the gravitational potential and non-planar effects do not significantly change the formation process (rotational fission) itself of asteroid binaries, but can slow down the overall evolutionary process, e.g. mutual escape occurs later in the simulations. \citet{2020Icar..34913849A} compare four different full two-body codes to determine the most optimal method to simulate the motion of the Didymos system. The two-body codes they consider model the asteroids as polyhedral or mascon shapes. They find that expanding the mutual potential up to order four is sufficient to describe the motion of the Didymos binary system.

However, an accurate shape model of an asteroid is often not available. Modeling an asteroid as a triaxial ellipsoid is commonly used to approximate the shape of the body to study the F2BP \citep{2009CeMDA.104..103S, 2011Icar..214..161J, 2016MNRAS.461.3982B, 2021CeMDA.133...35H, 2022alexpaper}, and the gravitational potential of such bodies can be expressed analytically \citep{MacMillan1930}.

Approximating the shape of asteroids as ellipsoids have previously been used to study the dynamics of post-fissioned asteroid systems \citep{2011Icar..214..161J, 2016MNRAS.461.3982B, 2022alexpaper}. In the rotational fission model, the initial separation between the two bodies is very small. In some cases, especially for non-planar cases, we might expect that a series approximation to the potential could cause erroneous values for both force and torque in the initial stages of the simulation, when the bounding spheres of the two bodies overlap and the power series diverges. Previous work on post-fissioned asteroid binary systems avoid this issue by imposing initial conditions which ensure that the bounding spheres do not intersect \citep{2011Icar..214..161J, 2017CeMDA.127..369H, 2016MNRAS.461.3982B}. 

We have in a series of papers \citep{2021CeMDA.133...27W, 2021CeMDA.133...35H, 2022alexpaper} investigated another approach to the F2BP, where the forces and torques (and mutual potential) are computed directly by integrating over the surface of one body in the gravitational field of the other \citep{Conway_2016}. For ellipsoidal bodies, the surface integration approach by \citet{Conway_2016} yield exact values for force, torque and mutual gravitational potential. \cite{2021CeMDA.133...27W} outline the surface integration and demonstrate the method in some torque-free planar cases of two spheroids and two disks. \citet{2021CeMDA.133...35H} extend this to non-planar cases, and also use it to study the dynamics of the 1999 KW4 system. While the surface integral method is exact for spheroids and triaxial ellipsoids, it can be somewhat computationally demanding as multiple double integrals must be evaluated at each time step. However, compared to evaluating triple integrals at each time step, it is very efficient. The surface integration method to compute the forces is exact for ellipsoidal bodies because it does not use series expansions, and it also produces exact results in cases when the bounding spheres of the two bodies overlap.

\citet{2017CeMDA.127..369H} used their method to compare the differences between different expansion orders for ellipsoidal shapes, and found that the discrepancy in the results becomes smaller with higher orders. However, no comparisons with a mathematically exact method have yet been performed. In this paper, we utilize our surface integration method to investigate the errors in force and torque produced by methods that calculate force and torque based from a series approximation of the mutual potential. We also explore what consequences these initial errors may have on the dynamical behaviour of a newly fissioned asteroid binary. For comparing with approximation-based methods, we have chosen to utilize the open source software ``General Use Binary Asteroid Simulator'' (\texttt{gubas})\footnote{Github repository: \url{https://github.com/meyeralexj/gubas}} developed by \citet{DAVIS2020113439}. \texttt{gubas} uses  the efficient algorithm based on recursive relations as described by \citet{2017CeMDA.127..369H}, and allows the user to choose the approximation order of the potential.

Section \ref{sec:Forces_and_torques} presents a brief review of the methods that are compared in this manuscript, and the technical details of the comparisons. In Sect. \ref{sec:Methods} we consider various configurations and study the difference in the values of the forces and torques using second and fourth order approximations and the surface integration method. Two test simulations are presented in Sect. \ref{sec:Sim_comparisons} to show the long-term consequences of using these two approaches on the prediction of the dynamic behavior of asteroid binaries. Sect. \ref{sec:CPU_comparison} compares the computational efficiency of the methods. A summary and discussion of our results are presented in Sect. \ref{sec:Discussion}. 

\section{Force, torque and mutual gravitational potential}
\label{sec:Forces_and_torques}
\subsection{Surface integration method}
In the surface integration method, the force $\mathbf{F}$ and torque $\mathbf{M}$ on an extended body with surface $S$,  surface normal $\mathbf{n}$ and density $\rho$, in the gravitational potential, $\Phi$, of another extended body are expressed by the following integrals
\begin{align}
\mathbf{F}(\tilde{\mathbf{r}}) &= \rho \iint\limits_S \Phi(\tilde{\mathbf{r}})\mathbf{n}dS \label{eq:Force_surface}
\\
\mathbf{M}(\tilde{\mathbf{r}}) &= -\rho \iint\limits_S \Phi(\tilde{\mathbf{r}})\mathbf{n}\times \tilde{\mathbf{r}} dS \label{eq:Torque_surface1} 
\end{align}
The mutual potential energy $U$ between the two bodies is 
also expressed via a surface integral
\begin{align}
U &= \frac{\rho}{3}\iint\limits_S\left[\tilde{\mathbf{r}}\Phi(\tilde{\mathbf{r}}) - \frac{1}{2}|\tilde{\mathbf{r}}|^2\mathbf{g}(\tilde{\mathbf{r}})\right]\cdot \mathbf{n}dS,
\label{eq:Potential_surface}
\end{align}
where $\mathbf{g}(\tilde{\mathbf{r}}) = \nabla \Phi$ is the gravitational field acting on the integrated body at a point described by the position vector $\tilde{\mathbf{r}}$. 
The position vector $\tilde{\mathbf{r}}$ is measured in the body-fixed frame of the body exerting the gravitational field \citep[see][for details]{2021CeMDA.133...27W, 2021CeMDA.133...35H}.

In this work, we assume that both bodies have uniform densities, and are triaxial ellipsoids. For a triaxial ellipsoid, $\Phi$ can be expressed analytically, hence Eqs. \eqref{eq:Force_surface}--\eqref{eq:Potential_surface} above give solutions to the force, torque and gravitational potential that are exact and not affected by truncation errors or other inaccuracies arising from using approximations. For the gravitational potential of a triaxial ellipsoid we use the expression derived by \citep{MacMillan1930}.
 
\subsection{Mutual gravitational potential expressed as power series}
Whereas in the surface integration method, the force and torque between two rigid bodies is calculated by integrating $\Phi$ over a surface, most other methods derive force and torque by first expanding $U$ in a series, and then differentiate $U$. We have chosen to compare with the output from the software \texttt{gubas} \citep{2020PSJ.....1...25D} where the mutual potential $U$ is expanded as  
\begin{align}
U = -G \sum_{n=0}^N\frac{1}{r^{n+1}}\tilde{U}_n
\label{eq:Hou_mutual_potential}
\end{align}
where $r$ is the separation between the mass centers of the two bodies, $G$ the gravitational constant, $N$ the truncation order of the potential, and $\tilde{U}$ contains the inertia integrals that are expanded with Legendre polynomials \citep{2017CeMDA.127..369H}. The force is computed as
\begin{align}
F_* = \frac{\partial U}{\partial (*)}, \text{ for }* = x, y, z
\label{eq:Force_gugbas}
\end{align}
and the torques as \citep{1995CeMDA..63....1M}
\begin{align}
\mathbf{M}_s'&=-\alpha_i\times \frac{\partial U}{\partial \alpha_i} -\beta_i \times \frac{\partial U}{\partial \beta_i} -\gamma_i\times\frac{\partial U}{\partial \gamma_i} \label{eq:TorqueA_gubas}\\
\mathbf{M}_p &= \mathbf{r}\times \frac{\partial U}{\partial \mathbf{r}}-\mathbf{M}_s'\label{eq:TorqueB_gubas}
\end{align}
where $\alpha_i, \beta_i, \gamma_i$ are the coordinate vectors of the secondary expressed in the body-fixed frame of the primary. The prime notation denotes the vector expressed in the body-fixed frame of the primary \citep{2017CeMDA.127..369H}. 

\begin{figure}
\centering
\includegraphics[width=\linewidth]{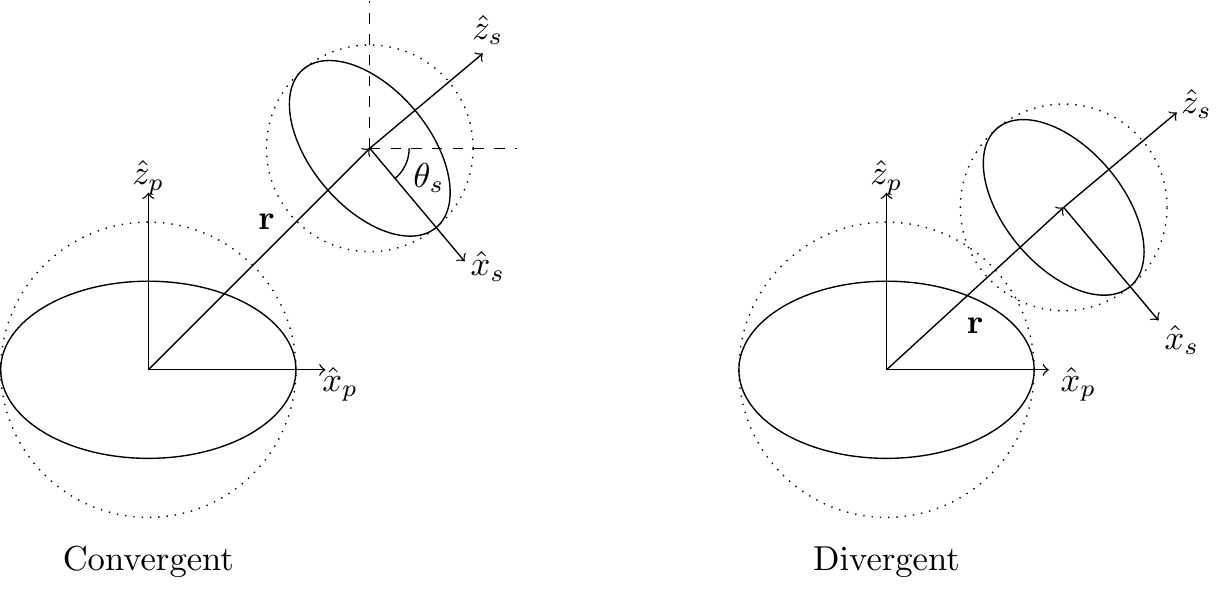}
\caption{Illustration of where the series expansion given by Eq.~\eqref{eq:Hou_mutual_potential} converges (\textit{left}) and diverges (\textit{right}) between two extended bodies. The dotted lines correspond to the bounding spheres around each respective body. In the figure, the secondary is rotated an angle $\theta_s$ around the $\hat{y}_s$-axis.  The hat variables denote body-fixed coordinates.}
\label{fig:Expansion_potential_convergence}
\end{figure}

The inertia integrals make use of Legendre polynomials to describe the mass distribution of the bodies, and therefore plays the same role as the spherical harmonics coefficients \citep{2008CeMDA.100..319T}. Similar to spherical harmonics, the power series described by Eq.~\eqref{eq:Hou_mutual_potential} converges in a certain region. \citet{2008CeMDA.100..319T} showed that the mutual potential, using this formulation, converges at every point outside the bounding spheres as long as the bounding spheres do not share any common points (see Fig \ref{fig:Expansion_potential_convergence} for an illustration). 

When using the mutual potential in Eq.~\eqref{eq:Hou_mutual_potential}, higher order terms with order $>N$ have been neglected, which leads to truncation errors. In summary, whereas obtaining force and torque from the truncated potential is computationally efficient, the disadvantages are a divergent potential for some configurations where the bodies are in close proximity, and truncation errors. The surface integral method does not encounter these disadvantages, but might be computationally more expensive, at least compared to lower-order potentials. This is because the computational efficiency decreases with a higher number of integral dimensions. Furthermore, the only integrals required to calculate Eq.~\eqref{eq:Hou_mutual_potential} are the inertia integrals, which are only solved once and can be re-used throughout the simulation \citep{2017CeMDA.127..369H}. 

Since we are able to fully solve the two-body problem for two triaxial ellipsoids in any non-overlapping configurations, without being affected by truncation errors in the mutual potential, this puts us in the position to investigate how truncation errors, and errors caused by divergence in the series approximation of $U$ might affect the ensuing dynamics of the binary system.

\section{Comparisons between the two methods}
\label{sec:Methods}
Thus there are two things we wish to investigate, truncation errors and errors related to overlapping bounding spheres. 

The first one we address by investigating the difference in force and torque between the two bodies for different positions of the secondary in the equatorial plane of the primary (for configurations where the bounding spheres do not overlap). The second one we address by investigating the difference in force and torque on the two bodies when the two bounding spheres overlap.

Finally, we run some longer simulations where
the equations of motion are solved in order to investigate
how any errors in force and torque made at the initial stages propagate in the ensuing dynamics. 

In order to extract values of the forces and torques from \texttt{gubas}, we made a slight modification to the software so that the computed forces and torques at the first time step are written to an external file. The forces and torques are also converted to standard SI units (m, s, and kg).

Whereas \texttt{gubas} uses relative coordinates for the position of the secondary relative to the primary, we use inertial frame coordinates for the positions of both bodies. Furthermore, in the formulation by \citet{2017CeMDA.127..369H}, the torque on the secondary is calculated in the body-fixed frame of the primary, while in our method the torque on the secondary is computed in its own body-fixed frame. In order to compare the torque on the secondary calculated in the two approaches, we therefore convert $\mathbf{M}_s'$ computed by \texttt{gubas} to the body-fixed frame of the secondary by the transformation
\begin{align}
\mathbf{M}_s = \mathcal{R}_s^T \mathcal{R}_p \mathbf{M}_s'
\end{align}
where $\mathcal{R}_p$ and $\mathcal{R}_s$ are the rotation matrices of the primary and secondary, respectively, and superscript $T$ denotes the transpose. 

Moreover, in order to compare velocities and positions from our method with that from \texttt{gubas}, we convert our positions and velocities to the body-fixed frame of the primary. The angular velocities from our code and from \texttt{gubas} are both expressed in the body-fixed frame of each respective body, hence there is no need for transformations of these. 

Throughout the manuscript, whenever we compute the relative difference between two vectors $\mathbf{v_i}$ and $\mathbf{v_j}$ (can be force, torque or angular/translational velocity) from model $i$ and $j$, respectively, we evaluate this as: 
\begin{align}
\delta v = \frac{|\mathbf{v}_i - \mathbf{v}_j|}{|\mathbf{v}_j|},
\label{eq:relerror}
\end{align}
where $|\, |$ denotes the norm.

For consistency, when we later (in Section~\ref{sec:Sim_comparisons}) solve the equations of motion, we use the same Runge-Kutta method with equal time steps, in both the surface integral method and in \texttt{gubas}. Any differences between the simulations should therefore not be affected by the choice of integration scheme.
 
The surface integration itself in our method is performed with the QAG adaptive integration algorithm from the QUADPACK implementation in the GNU scientific library \citep{galassi2002gnu}. The integration order can be selected from one to six, and using higher orders increases the accuracy while reducing the computational efficiency. Unless otherwise specified, we use the sixth-order QAG integrator\footnote{Using an order four QAG integrator can reduce the computation time by a factor of two.}.

\subsection{Effect of truncation errors in mutual potential on the force and torque}
\label{sec:Force_comparisons}
For the first experiment, we assume that the binary consists of a primary with semiaxes $(a_p, b_p, c_p) =$ (400 m, 390 m, 350 m) and a secondary
with $(a_s, b_s, c_s) =$ (100 m, 90 m, 80 m), and both have densities of $\rho = 2000$ kg m$^{-3}$, corresponding to a mass ratio of $q=m_s/m_p=0.013$. A number of observed binaries have estimated mass ratios close to this value \citep{2016Icar..267..267P, 2020Icar..34813777N}.

\begin{figure}
\centering
\includegraphics[width=\linewidth]{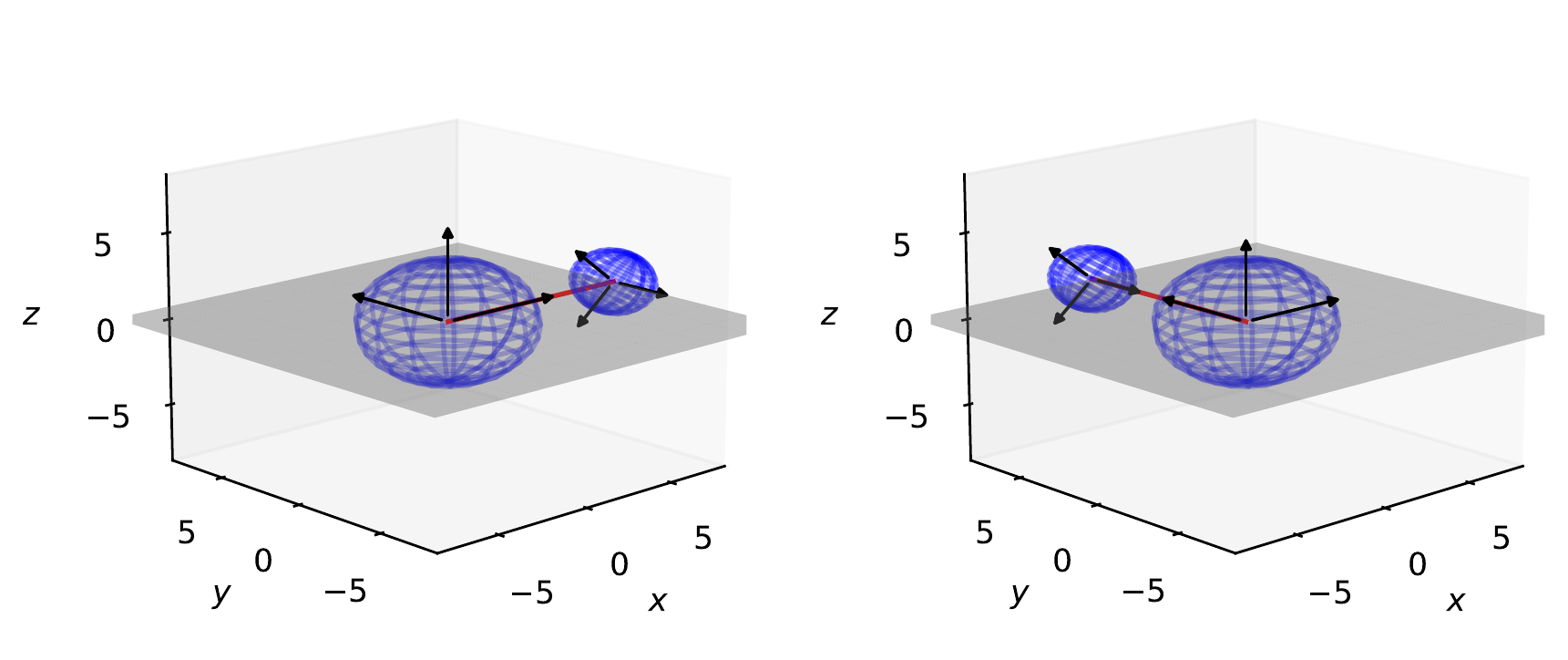}
\caption{Illustration of how the secondary is placed in the $xy$-plane of the primary. The red line corresponds to the separation vector $\mathbf{r}$ between the centroids. In the \textit{left panel}, $\mathbf{r}$ is parallel to one of the principal axes of the primary, whereas in the \textit{right panel}, it is parallel to a principal axis of both bodies. The axes are given in dimensionless quantities.}
\label{fig:moving_ellipsoid_plane}
\end{figure}

The secondary is first rotated an angle $\theta_s=45^\circ$ about its $y$-axis (see Fig.~\ref{fig:Expansion_potential_convergence}), and then placed with its centroid in the equatorial plane ($xy$-plane) of the primary (see Fig \ref{fig:moving_ellipsoid_plane}). In this manner, the configuration is made non-planar. The secondary is thereafter placed at a number of different positions in the $xy$-plane of the primary, so that the distance between them varies from a minimum value up to a maximum value of five primary radii (we take $a_p$ to be the primary radius). Asteroid binary observations by \citet{2016Icar..267..267P} show that the orbits of the secondaries have semimajor axes between three to seven times the primary radius, hence our chosen range corresponds to common distances found in nature. The minimum distance at which we place the secondary corresponds to the distance when the two bounding spheres of the bodies start to intersect. This is to ensure that the mutual potential described by Eq.~\eqref{eq:Hou_mutual_potential} converges.

For each position of the secondary in the equatorial plane of the primary, we compute the forces and the torques on both the primary and the secondary using the surface integral method, and using a second and fourth order mutual potential with \texttt{gubas}. In this manner, can can study how the errors in the force and torques change with increasing separation between the bodies and with the order of the potential. In our calculations, we have rounded force and torque components with magnitudes smaller than $10^{-16}$ off to zero. 

\begin{figure}
\centering
\includegraphics[width=\linewidth]{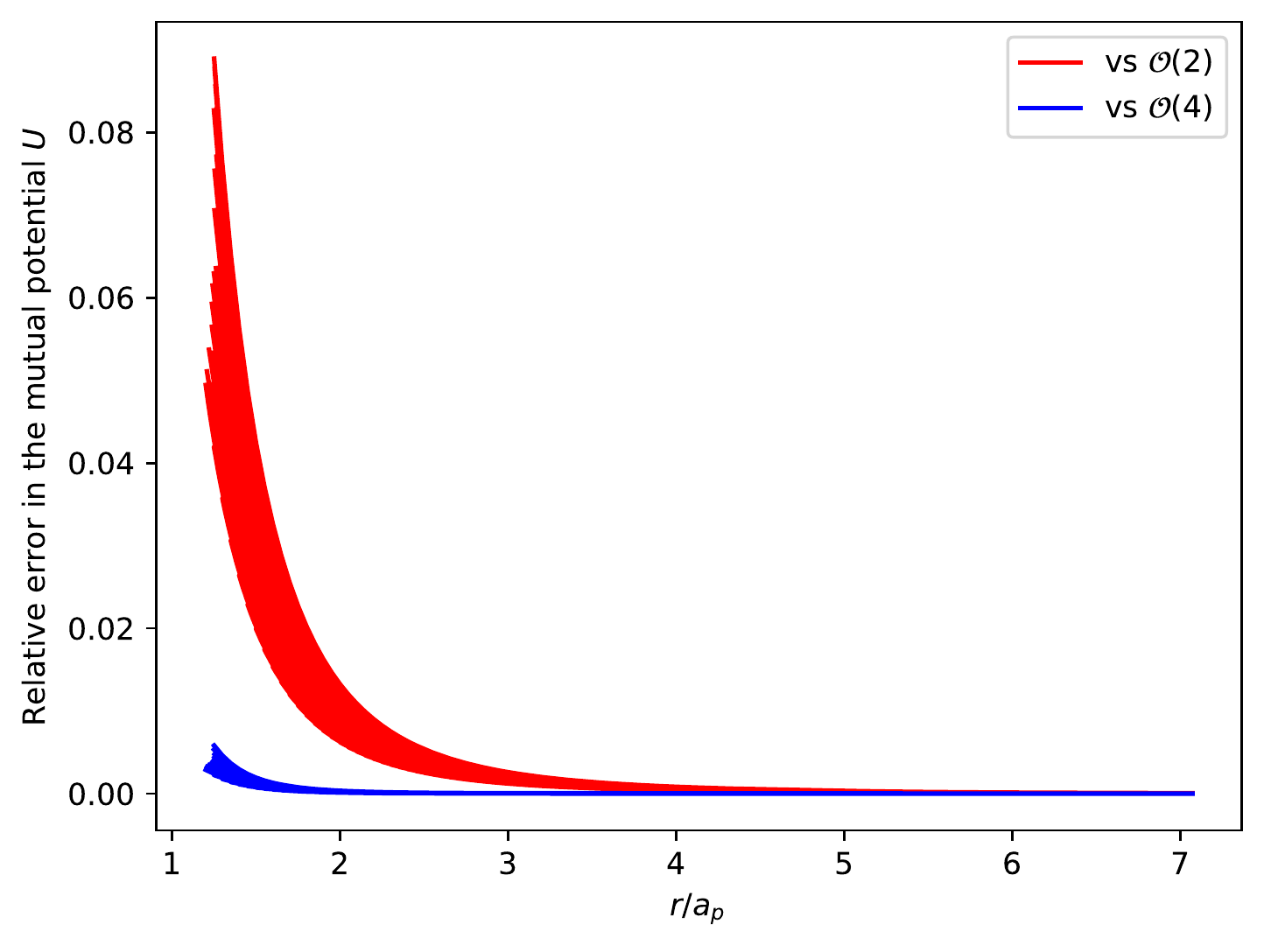}
\caption{Relative error (in percentage) in the mutual potential $U$ from the second (red line) and fourth order (blue line) potentials compared to the surface integration method, as functions of the separation in primary radii $a_p$.}
\label{fig:mutualpotential_reldiff_contour}
\end{figure}

We first consider the mutual potential differ at various separations between the methods, which is shown in Fig.~\ref{fig:mutualpotential_reldiff_contour}. The errors in the mutual potential are smaller than 0.09\% and 0.006\% for the second and and fourth order potentials respectively. The largest error, for this particular scenario, does not occur at the minimum separation, but takes place at approximately 1.25 primary radii. Our results are similar to the results of \citet{2015MNRAS.450.3742C}, who compared their method with the polyhedron approach by \citet{2012Geop...77F...1T}, where they found that the largest discrepancy in the gravitational potential occurred near the edges of the asteroid in which the distance to the body's center of mass is the largest.

\begin{figure*}
\centering
\includegraphics[width=\linewidth]{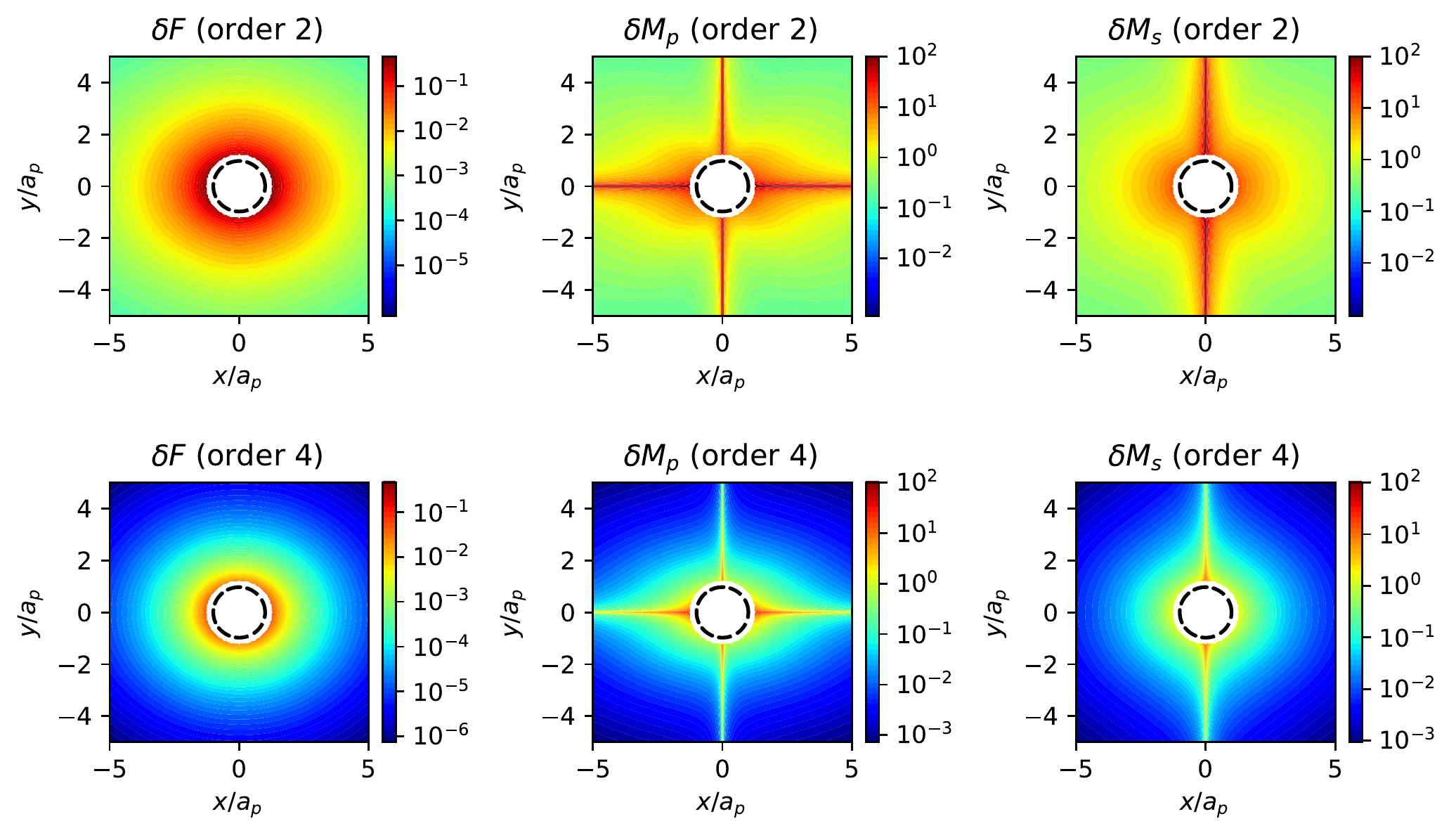}
\caption{Relative error (in percentage) in force and torques arising from using second (\textit{top}) and fourth order (\textit{bottom}) mutual potentials. The two \textit{left-most panels} show the force, and the \textit{middle} and \textit{right-most panels} show the torque on the primary and secondary, respectively. Each position in the $xy$-plane corresponds to the position of the secondary in the $xy$-plane of the primary. The units of the axes are units of primary radius $a_p$. The black dashed circles indicate the circumference of the primary, and the white region in the center corresponds to the region where the bounding spheres intersect. The color scaling is logarithmic and the color bars show numbers in percentage. The pair-wise comparisons between second and fourth order potentials share the same color scaling.}
\label{fig:force_torque_reldiff_contour}
\end{figure*}

The results of the calculations of the forces and torques are shown in Fig.~\ref{fig:force_torque_reldiff_contour} where the relative difference between the surface integration method and the two expansion approaches is shown as a color scale in the $xy$-plane of the primary. The panels to the left show the relative difference in force, while the middle and right panels show the relative difference in the torque on the primary and secondary, respectively.

In the left-most panels it can be seen that the relative errors in the force are largest when the bodies are close: $\sim 0.4$\% for the second order approximation, and roughly an order of magnitude smaller for the fourth order approximation. As the separation becomes larger, the errors decrease, and become negligible ($<0.001$\%), consistent with what is to be expected. For the fourth order approximation, the error in the force has already dropped to $0.001$\% when the distance between the bodies is 2-3$a_p$. Overall, we see that the relative error in force from the fourth order potential is roughly an order of magnitude smaller than that from the second order potential, as a fourth order expansion is closer to the exact solution with smaller truncation errors.

We now consider the relative error in the torque on the primary as shown in the middle panels. When the secondary's centroid is placed either on the $x$- or $y$-axis of the primary, the error in the torque on the primary, $\delta M_p$, is 100\% for the second order potential, regardless of the separation between the two bodies. This happens because in this configuration the vector between the centroids of the two bodies, $\mathbf{r}$, is parallel with the principal axes of the primary. In these configurations, the second order approximation yields a vanishing torque on the primary \citep{kane1983spacecraft, 2012JCoPh.231.7237P}. The zero torque from the second order approximation is  however unrealistic in this case, as the torque calculation from both the surface integration method and fourth order potential indicates that non-zero torques are experienced by the primary. 

For the torque on the secondary, as is shown in the right-most panels in Fig.~\ref{fig:force_torque_reldiff_contour}, the 100\% error, from the second order approximation, occurs only when it is placed with its centroid on the $y$-axis of the primary. Similar to the torque on the primary, this happens because $\mathbf{r}$ is parallel with a principal axis (in this case, the intermediate-axis) of the secondary (see right panel of Fig.~\ref{fig:moving_ellipsoid_plane}). At other regions in the $xy$-plane where $\mathbf{r}$ is not parallel with any one of the principal axes, the errors in $\mathbf{M_p}$, when using the second order potential, range between $\sim 2$\% and $\sim 10$\% when the bodies are close, and drops to $\sim 1$\% at larger distances. The error in $\mathbf{M_s}$, on the other hand, is $\sim 10$ \% at the smallest separation and $\sim 3$\% at the largest distances. Furthermore, similar to the force, the relative errors in $\mathbf{M_p}$ and $\mathbf{M_s}$ using the fourth order potential are roughly an order of magnitude smaller than when using the second order potential. 

It is clear from Fig.~\ref{fig:force_torque_reldiff_contour} and from the discussion above that the relative error in the torques is larger than in the force. This is also seen in other work that involves expansions to study electrostatic forces \citep{2012JCoPh.231.7237P, poursina2020electrostatic}. We therefore argue that using approximations to the mutual potential may have a larger effect on the rotational motion of the bodies than on the translational motion.

We briefly investigate how the mass ratio of the system may affect the differences in the computed forces and torques. The semiaxes of the secondary are changed to $(a_s, b_s, c_s) =$ (250 m, 240 m, 230 m), while keeping the semiaxes of the primary and the densities of the bodies the same, which corresponds to a mass ratio of $q=0.25$. The resulting errors in forces and torques are slightly lower, but similar, to that of Fig.~\ref{fig:force_torque_reldiff_contour}. However, the decrease in the errors are less than one percent. This suggests that the mass ratio of the system should not significantly affect the computed forces and torques, provided that the bodies are sufficiently far apart.

\subsection{Primary and secondary with overlapping bounding spheres}
\label{subsec:Surface_touch}
In this section, we investigate situations where the bounding spheres of the bodies overlap and the mutual potential described by Eq.~\eqref{eq:Hou_mutual_potential} no longer converges \citep{2008CeMDA.100..319T}. This happens when the surface of the secondary is allowed to almost touch the surface of the primary. This type of configuration is particularly relevant for newly fissioned asteroid systems, as the bodies are initially very close. 

\begin{figure*}
\centering
\includegraphics[width=0.7\linewidth]{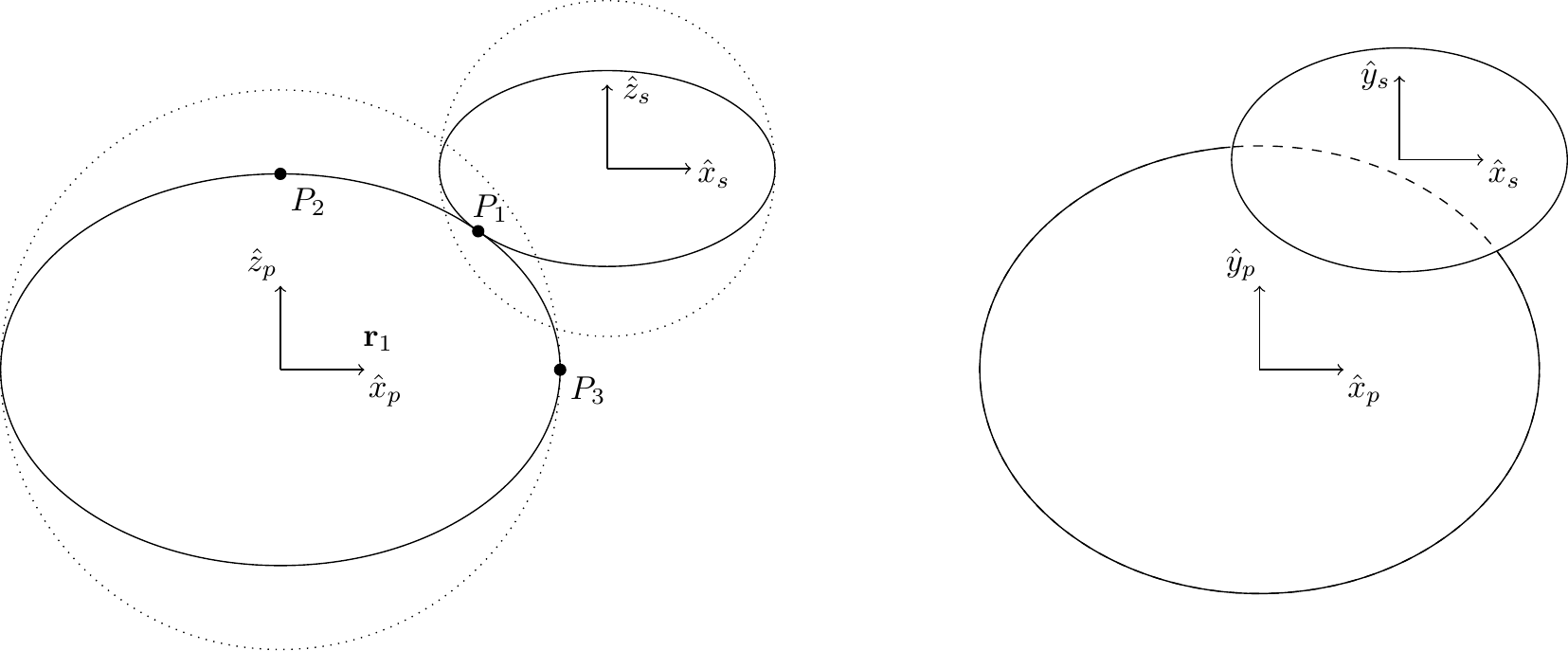}
\caption{Example configurations where the surfaces of the ellipsoids are almost touching. The \textit{left panel} shows a view of the $xz$-plane, where the bodies are almost touching at point $P_1$. The bounding spheres around each body are marked with dotted lines, and are seen to overlap. Point $P_2$ is at the pole of the primary, and $P_3$ at the equator. Each connection point $P_i$ corresponds to a position vector $\mathbf{r}_i$ between the centroids of the bodies. The \textit{right panel} shows a view from  above of the $xy$-plane.}
\label{fig:Case2_config_illu}
\end{figure*}

Figure \ref{fig:Case2_config_illu} shows two examples from two different viewing angles of the configurations we investigate in this section. The position of the secondary is such that the separation between the ellipsoids is the shortest while ensuring that the surfaces of the bodies do not overlap. The bodies {\em nearly} touch, i.e.\ there is no normal force involved in our calculations. We choose a number of different positions of the secondary such that the point $P$ is distributed over the entire upper half of the surface of the primary (because of symmetry, we only consider surface connection points on the upper half of the primary). Contrary to what was done in the previous section, we now keep all three axes of each of the body-fixed coordinate systems parallel.

We compute force and torques as in the previous section, but for three different shapes of the primary. The long semiaxis of the primary is fixed at $a_p = 400$ m, and three values of the axis ratios $a_p/b_p$ and $a_p/c_p$ are chosen. The three different shape models of the primary are listed in Tab. \ref{tab:Shape_ratio_surface_touch}, one is a spheroid (Model 1) and two of them are rather elongated (Models 2 and 3). The secondary is kept at the same size and shape as in the previous section. 

\begin{table}
\centering
\begin{tabular}{|l|c|c|c|}
\hline 
Model & $a_p/b_p$ & $a_p/c_p$ & $q$ \\
\hline
Model 1 & 1.000 & 1.067 & 0.012\\
Model 2 & 1.231 & 2.000 & 0.028\\
Model 3 & 1.455 & 4.000 & 0.065\\
\hline
\end{tabular}
\caption{Parameters used for the three models chosen for the tests in Sect. \ref{subsec:Surface_touch}. The second and third columns show the axis ratios of the primary, while the fourth column shows the mass ratio, $q=m_s/m_p$, of the system. Model 1 is a spheroid, and Models 2 and 3 are elongated ellipsoids. The long semiaxis of the primary is fixed at $a_p=400$ m.}
\label{tab:Shape_ratio_surface_touch}
\end{table}

\begin{figure*}
\centering
\includegraphics[width=\linewidth]{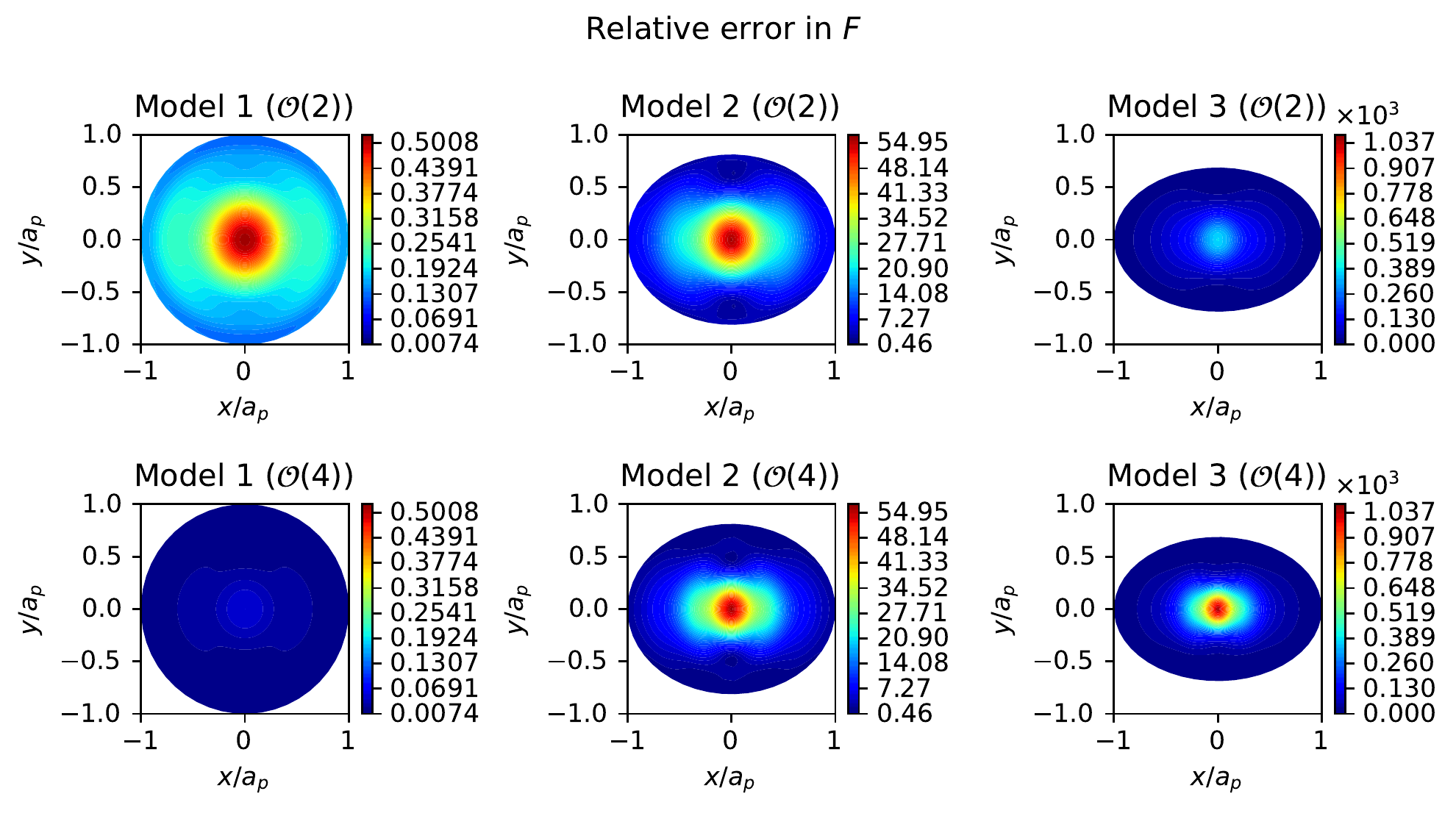}
\caption{Percentage difference between force $F$ from \texttt{gubas} using second and fourth order potentials, relative to force from surface integration method. The $xy$-plane shows the surface of the primary viewed from the top, along the $z$-axis. Each point in the plane corresponds to a surface connection point on the primary,
i.e.\ a new position of the secondary. The \textit{top} and \textit{bottom panels} are for second and fourth order approximations, respectively, and the panels show results for three different models of the primary, see Table \ref{tab:Shape_ratio_surface_touch}. The pair-wise comparisons between second and fourth order potentials share the same color scaling.}
\label{fig:Error_surface_touch}
\end{figure*}

\begin{figure*}
\centering
\includegraphics[width=\linewidth]{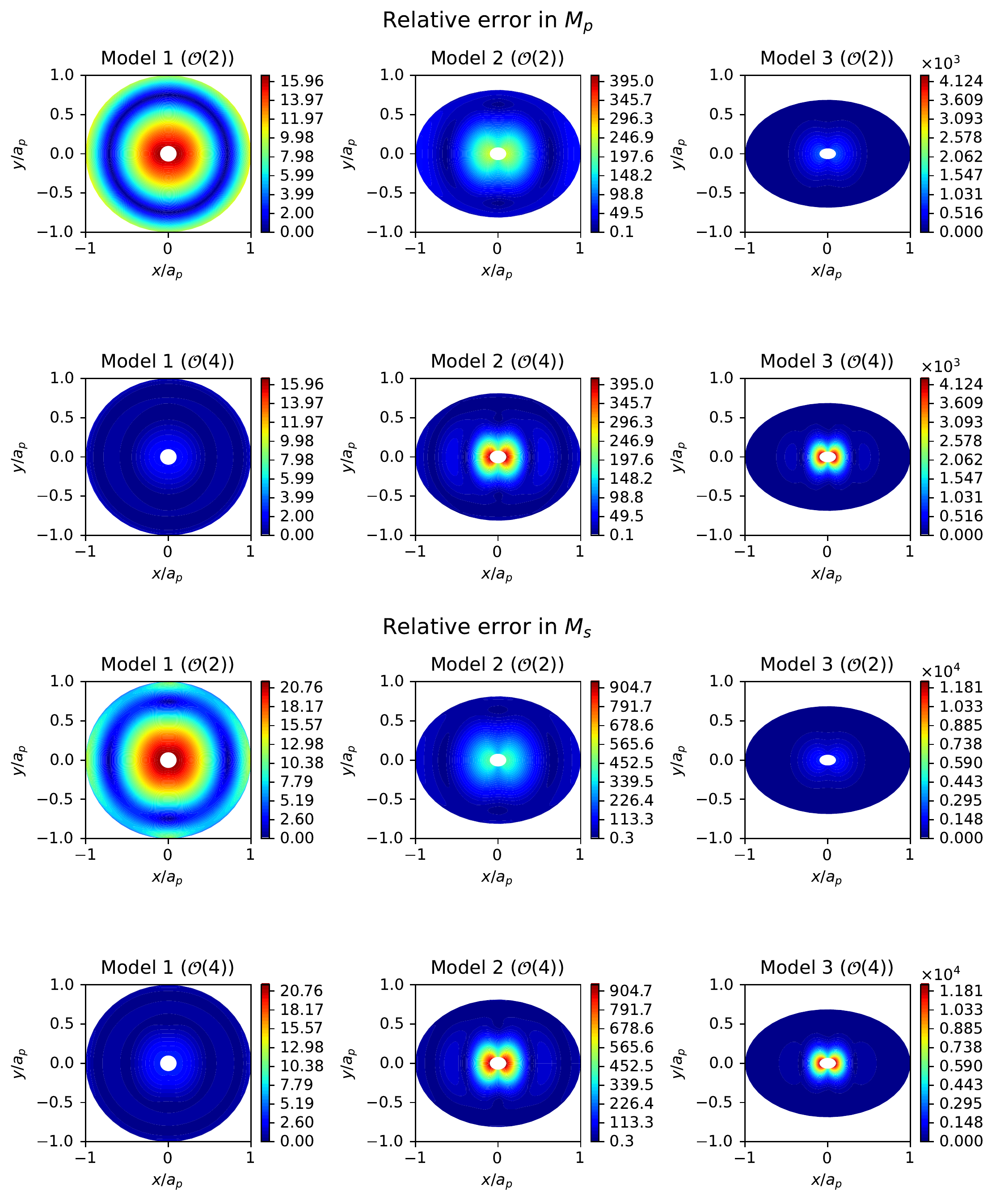}
\caption{Same as Fig. \ref{fig:Error_surface_touch}, but now showing the error in the computed torques on both bodies. The \textit{first} and \textit{second rows} show the relative differences in the torque on the primary $M_p$, while the \textit{third} and \textit{fourth rows} show the difference in the torque on the secondary $M_s$. Within the white region, the torques acted on the bodies are zero, and we have chosen to exclude this region from the plot as it is difficult to show relative error in $M$ when $M$ is close to zero.}
\label{fig:Error_surface_touch_M}
\end{figure*}

The results of the comparison between the output from \texttt{gubas} and the output from the surface integration method are shown in Fig.~\ref{fig:Error_surface_touch} and Fig.~\ref{fig:Error_surface_touch_M}, again as colored contour plots. All panels in the figures show the surface of the primary ellipsoid viewed from above, along the $z$-axis, and each point in the $xy$-plane represents the connection point $P$ on the surface of the primary as illustrated in Fig.~\ref{fig:Case2_config_illu}.

Fig.~\ref{fig:Error_surface_touch} shows the relative error in the force that arises when using second or fourth order potentials. The error increases as the secondary is moved closer to the pole of the primary (point $P_2$ in the left panel of Fig.~\ref{fig:Case2_config_illu}). At this location the distance between the centroids is at the minimum, and the error in the force is the largest. For the spheroidal primary (Model 1), the error is rather small for both the second and fourth order approximations ($<0.5$\%), whereas for the elongated models it ranges from $\sim50$\% to above $1000$\%. 

For the more elongated model (Model 3), the errors from the fourth order approximation become larger than that from the second order. As the separation between the mass centers $r$ becomes smaller, the bounding spheres will overlap more, causing a larger error in the mutual potential when it is expanded through power series. Furthermore, the forces between two extended bodies, obtained from expanding inertia integrals up to order $N$, scale as $(a_p/r)^N$ \citep{kane1983spacecraft}. For nearly all configurations we have considered here, we have that $r < a_p$. The higher order gravity terms will therefore result in values larger than the lower order terms, thus inflating the values of the computed forces. The forces from the expansion method therefore become greater than the values obtained with the surface integration method.

The error in the torque on the primary arising from using second and fourth order potentials is shown in the first and second rows of Fig.~\ref{fig:Error_surface_touch_M}.  Similar to the force calculations, the error is larger for the most elongated primary, particularly near the pole, regardless of whether the expansion order is two or four. As for the force calculations, when the primary becomes more elongated, the relative error using the fourth order approximation becomes greater than when using the second order approximation. This is again because the separation $r$ becomes smaller, and higher order gravity terms inflate the computed torques.

There are five surface connection points on the primary where the torque is zero. These are located at $(x=\pm a_p, y=0)$ and at $(x=0, y=\pm b_p)$ (along the equator) and at $(x=0,y=0)$ (the pole). Only the point in the pole is included in Fig.~\ref{fig:Error_surface_touch_M} (white region). Away from the region around the pole of the primary, Model 1 yields relative errors in $M_p$ up to 16\% with the second order approximation, while the error from using fourth order approximations is smaller by one magnitude. Hence for a spheroidal primary, both the fourth and the second order method give relatively good approximations of the torque, despite the bounding spheres overlapping.

The results for the torque on the secondary are shown in the third and fourth rows of Fig.~\ref{fig:Error_surface_touch_M}. The relative error behaves in much the same way as for the primary, but is significantly larger in magnitude for model 3, as the errors can reach as high as $10^4$\% for the fourth order approximation.

In the next section, we investigate what consequences these initial errors in force and torque may have in dynamical simulations. 

\section{Dynamical simulations of binaries}
\label{sec:Sim_comparisons}
We run two different simulations in this section, in the first one the two asteroids are spaced relatively far apart, and in the other they resemble a contact binary that just separated into two components via rotational fission. 

For solving the equations of motion, we use the standard fourth order Runge-Kutta method. For the surface integration method, the rotational motion is described using Euler parameters (i.e. quaternions), and the equations of motion of the bodies are described in \citet{2021CeMDA.133...35H}.

\subsection{Scenario 1: A binary with moderate separation}
\label{subsec:Simple_demo}

In the first simulation, the asteroids orbit each other at a distance of  three to four times the primary radius, thus resembling some observed binaries \citep{2016Icar..267..267P}. The semiaxes and densities of the two bodies are the same as in Sect.~\ref{sec:Force_comparisons}. We place the secondary initially at the position $\mathbf{r} =  [1800, 0, 5]$ m relative to the primary, and give it an initial velocity $\mathbf{v} =  [0, 0.12, 0]$ m s$^{-1}$. The primary has an initial angular velocity of $\boldsymbol{\omega}_p = [0, 0, 10^{-4}]$ rad s$^{-1}$, whereas the secondary has zero initial angular velocity. The bodies are also placed initially such that their body-fixed axes are parallel.  The integration time is 100 days, with a fixed time step of five minutes.

\begin{figure}
\centering
\includegraphics[width=\linewidth]{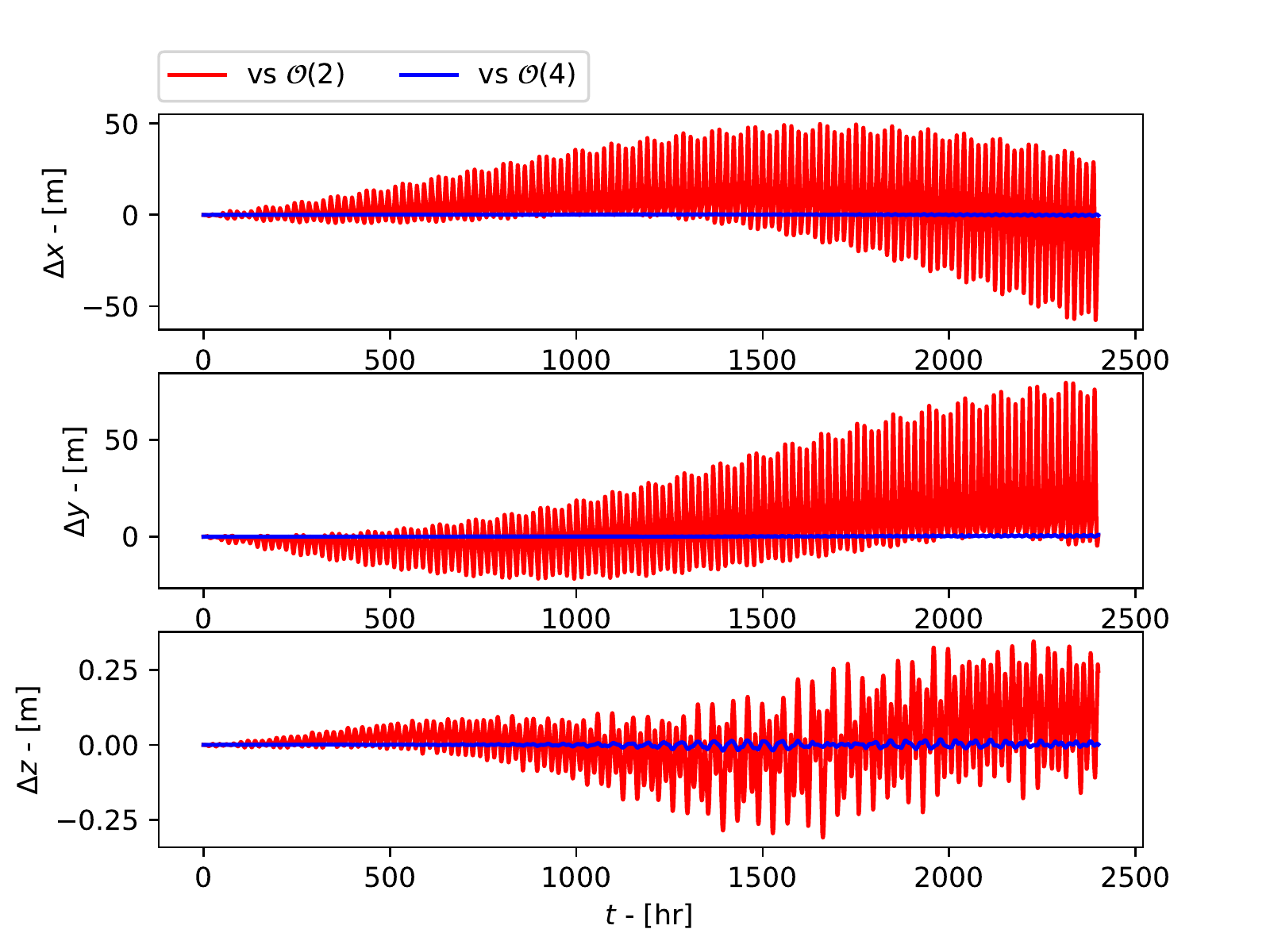}
\caption{For scenario 1: The difference between between \texttt{gubas} and our method for the position of the secondary relative to the primary. The \textit{top}, \textit{middle} and \textit{bottom} shows the difference in the $x$, $y$ and $z$ components respectively. The red line is the difference between \texttt{gubas} second order approximation and the surface integration method, whereas the blue line shows the difference between \texttt{gubas} fourth order approximation and the surface integration method.}
\label{fig:Basic_demo_position_diff}
\end{figure}

Figure \ref{fig:Basic_demo_position_diff} shows the difference between the output from \texttt{gubas} and the surface integration method for the $x$, $y$ and $z$-components of the position of the secondary. In the second order approximation, the $x$ and $y$ position of the secondary fails by approximately $\pm50$ m at the most ($\sim 3$\% of the distance between the primary and secondary), and for the $z$-component with $\pm0.25$ m at the most. The position calculated with the fourth order approximation in \texttt{gubas}, is indeed a very good approximation, as shown by the blue line in Fig.~\ref{fig:Basic_demo_position_diff}, where the differences are smaller than $\sim 1$ m for all three components. This agrees with the findings in Sect. \ref{sec:Force_comparisons}, where the errors from the fourth order potential is an order magnitude smaller than the second order approximation. 

\citet{2017CeMDA.127..369H} compare how the $x$-position of the secondary deviates between different orders of the potential, with an initial separation of $3.6$ times the primary radius. The deviation of the $x$-position, between the second and fourth order potential, surpassed over $1000$ m after $\sim 130$ hours. In our simulations, after 130 hours, the deviation in the $x$-position is $\sim 1.4$ m and $\sim 10^{-3}$ m for the second and fourth order approximations respectively. Our comparison between the surface integration scheme and the second order potential is similar to the order ten and order eight comparison performed by \citet{2017CeMDA.127..369H}, where the $x$-position deviated by $\sim 5$ m after 130 hours. The system considered by \citet{2017CeMDA.127..369H} has a mass ratio of $q = 0.512$. However, as discussed in Sect.~\ref{sec:Force_comparisons}, the mass ratio of system should not significantly alter the differences in the computed forces, as long as the bodies are sufficiently far apart.

\begin{figure}
\centering
\includegraphics[width=\linewidth]{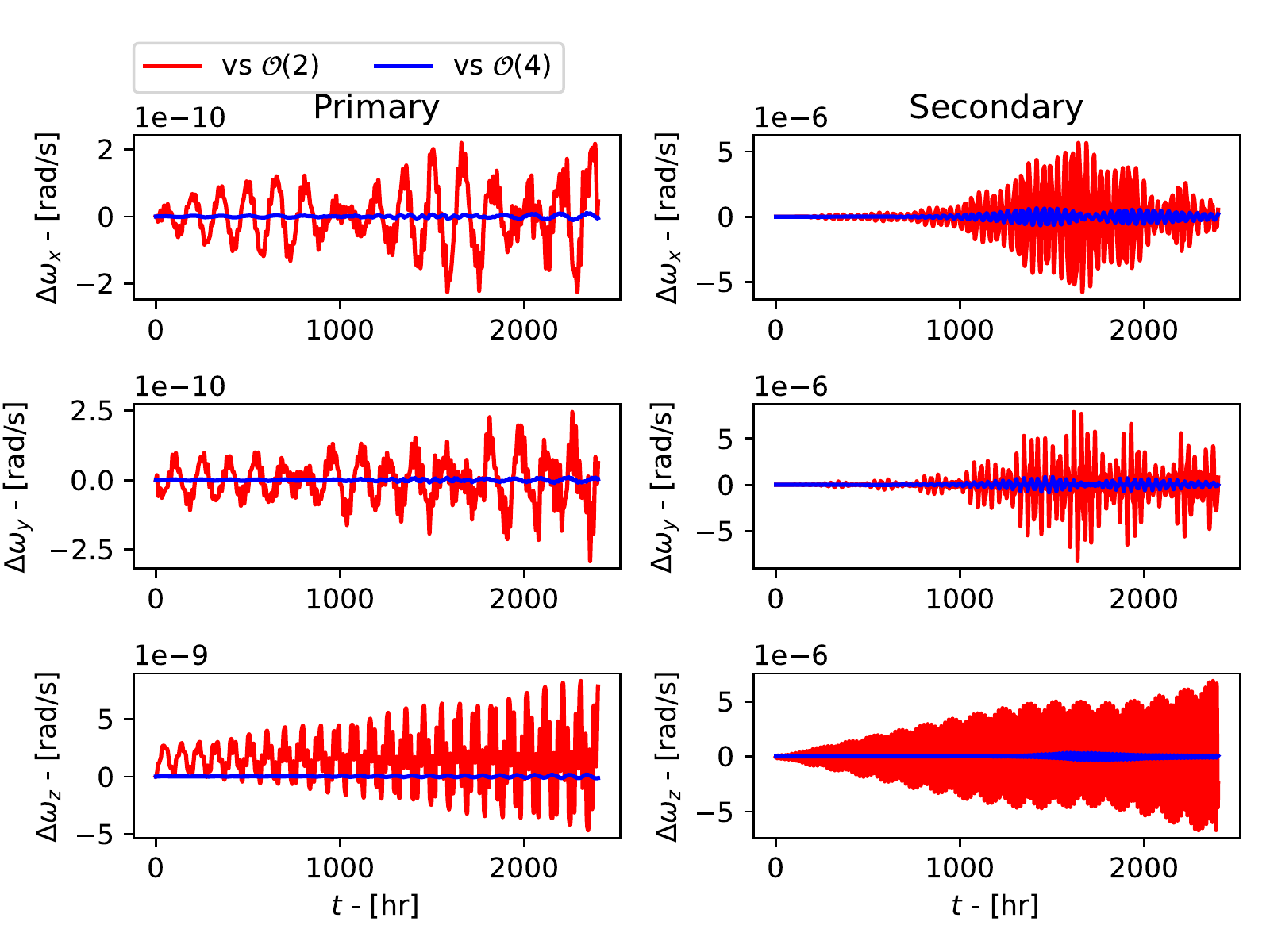}
\caption{Same as Fig.~\ref{fig:Basic_demo_position_diff}, but for the components of the angular velocity. The \textit{left panels} show the components for the primary, and the \textit{right panels} show the components for the secondary.}
\label{fig:Basic_demo_angvel_diff}
\end{figure}

In Fig.~\ref{fig:Basic_demo_angvel_diff} we plot the differences in the components of the 
angular velocities of both the primary and secondary throughout the simulation. The differences between the second order approximation and the surface integration method are of order $10^{-10}$ rad/s and $10^{-6}$ rad/s for the primary and secondary, respectively, 
and the difference is reduced by a factor of ten when the fourth order potential is used. 

\begin{figure}
\centering
\includegraphics[width=\linewidth]{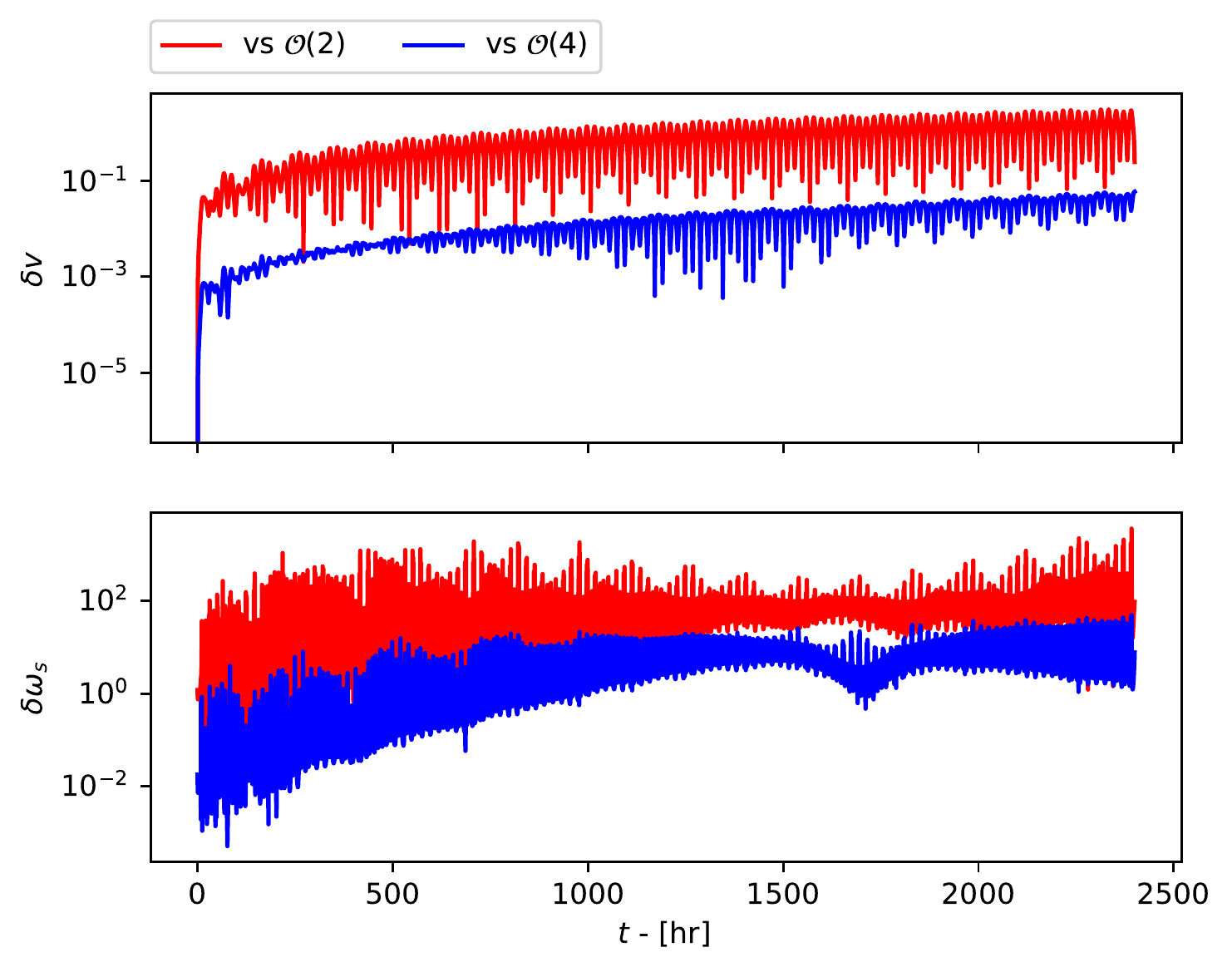}
\caption{Relative difference (in percentage) between the models for the 
speed (\textit{top}) and the angular speed (\textit{bottom}) of the secondary. The red line shows the second order potential approximation relative to the surface integral method, and the blue line the fourth order method relative to the surface integration method.}
\label{fig:Basic_demo_reldiffs}
\end{figure}

As the results from Sect.~\ref{sec:Methods} indicate that the choice of approximation order for the potential affects the rotational motion more significantly than the translational motion, we wish to compare differences in translational and rotational velocity. This is shown in Fig.~\ref{fig:Basic_demo_reldiffs} where we have plotted the difference in the velocity and the angular velocity of the secondary, as calculated from both the second and fourth order potential relative to the surface integration method. For the second order potential, the relative difference in the translational velocity is under $\sim 3$\%, while the relative difference in the angular velocity averages at $\sim 70$\%. The error in the rotation period of the secondary, computed as $T_s = 2\pi/|\boldsymbol{\omega}_s|$, also averages at roughly 70\%. This showcases that using the surface integration scheme to determine the motion of asteroids, in which the results are exact for ellipsoidal shapes, is more important to correctly predict rotational motion. 

The Double Asteroid Redirection Test (DART) is a NASA mission that aims to demonstrate how a kinetic impactor can be used to redirect the orbits of objects that may potentially collide with Earth \citep{2021PSJ.....2..173R}. One of the observable quantities after the impact is the orbital period of the secondary, and may fluctuate over time scales from days to months depending on the shape of the target body and the momentum transfer enhancement factor \citep{2022PSJ.....3..157R}.  It is therefore interesting to briefly check whether the approximation order of the potential significantly influences the period of the secondary in a binary system. In doing this, we find (for the assumed binary in this section) a relative error of $<0.1$\% in the period from using the second order potential, and for the fourth order potential a relative error in the period of $<0.001$\%. The former corresponds to a difference in $\sim 3.6$ seconds in the orbital period, while the latter a difference of $\sim 0.4$ seconds.

In summary, for the assumed binary in this section, a variation of $\sim \pm10$ m in the position of the secondary, $\pm10^{-6}$ rad/s in the angular velocities, and $<0.1$\% in the seconday's orbital period are small enough to be negligible for the overall orbit. Hence, provided the components are far enough apart, using the fourth order potential is sufficient to describe the dynamics of asteroid systems, such as the Didymos binary system \citep[see also][]{2020Icar..34913849A}. However, these results are based on asteroids with perfect ellipsoidal shapes, while real asteroids are better described with e.g. polyhedral shapes.

\subsection{Scenario 2: A fissioned contact binary}
\label{subsec:Rotfission_sim}
In this section, we simulate an asteroid binary after a contact binary has separated into two components due to rotational fission. The bodies in this system are initially very close so that their bounding spheres overlap in the initial stages of the simulation. We compare the output from simulations using the approximative method from \texttt{gubas} with that from our surface integration method. 

\begin{figure}
\centering
\includegraphics[width=\linewidth]{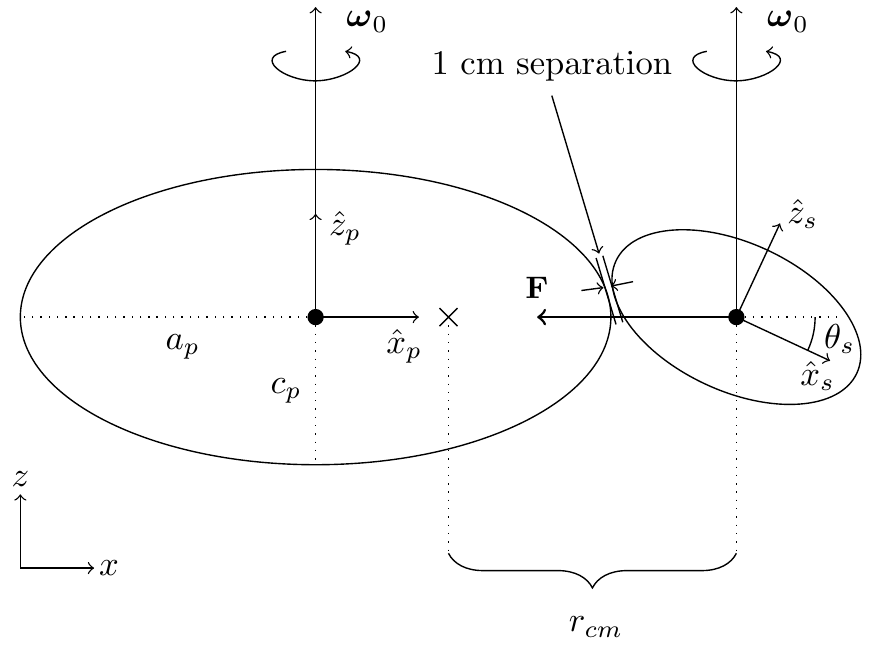}
\caption{Illustration of the initial configuration for the rotational fission scenario described in Sect. \ref{subsec:Rotfission_sim}.}
\label{fig:RotFission_initial_illustration}
\end{figure}

The initial conditions are the same as those of \citet{2022alexpaper}, where the secondary is initially rotated by an angle $\theta_s = 5^\circ$ about its body-fixed $y$-axis and the surface-to-surface distance between the primary and secondary is 1 cm (see Fig.~\ref{fig:RotFission_initial_illustration}). The semiaxes of the primary are $(a_p, b_p, c_p) = (1000, 700, 650)$ m, and for the secondary $(a_s,b_s,c_s) = (699,469,435)$ m, chosen so that the mass ratio is $q=0.3$. This mass ratio is large enough to yield a negative total energy for the system so that the components do not undergo mutual escape. The integration time is one year with time steps of five minutes.

Given the selected semiaxes of the two bodies and the required surface-to-surface distance of 1 cm, the initial position of the secondary becomes $(x,y,x) = (1667.2,0,0)$ m, while the primary is located at the origin. By equating the centrifugal force with the gravitational attraction in this configuration, we find the initial angular velocity $\omega_0$ that the system must have in order to undergo rotational fission \citep[for details see][]{2022alexpaper}
\begin{align}
    \omega_0 = \beta \sqrt{\frac{F}{m_s r_{cm}}},
\end{align}
where $F$ is the magnitude of the gravitational force, $m_s$ the mass of the secondary, $r_{cm}$ the distance between centroid of the secondary and the center of mass of the system (see Fig.~\ref{fig:Case2_config_illu}) and $\beta$ a cohesion factor. Following \citet{2022alexpaper}, we use $\beta = 1.01$. The gravitational forces used to determine $\omega_0$ are obtained by measuring $\mathbf{F}$ for the given initial configuration for all three models (similarly to what is done in Sect. \ref{sec:Methods}). For the chosen configuration we get a numerical value of $\omega_0 = 2.99 \cdot 10^{-4}$ rad/s (spin period of 5.84 hr) from the surface integration method. For the second and fourth order approximation methods, we find  $\omega_0 = 2.92\cdot 10^{-4}$ rad/s and $2.97 \cdot 10^{-4}$ rad/s, respectively (corresponding to spin periods of 5.98 hrs and 5.88 hrs). Conservation of angular momentum thereafter gives the initial translational velocities of the components \citep[for details, see][]{2022alexpaper}. Thus the fission
limit $\omega_0$ is slightly different in the three cases because it ultimately depends on the mutual gravitational potential. 

\begin{figure}
\centering
\includegraphics[width=\linewidth]{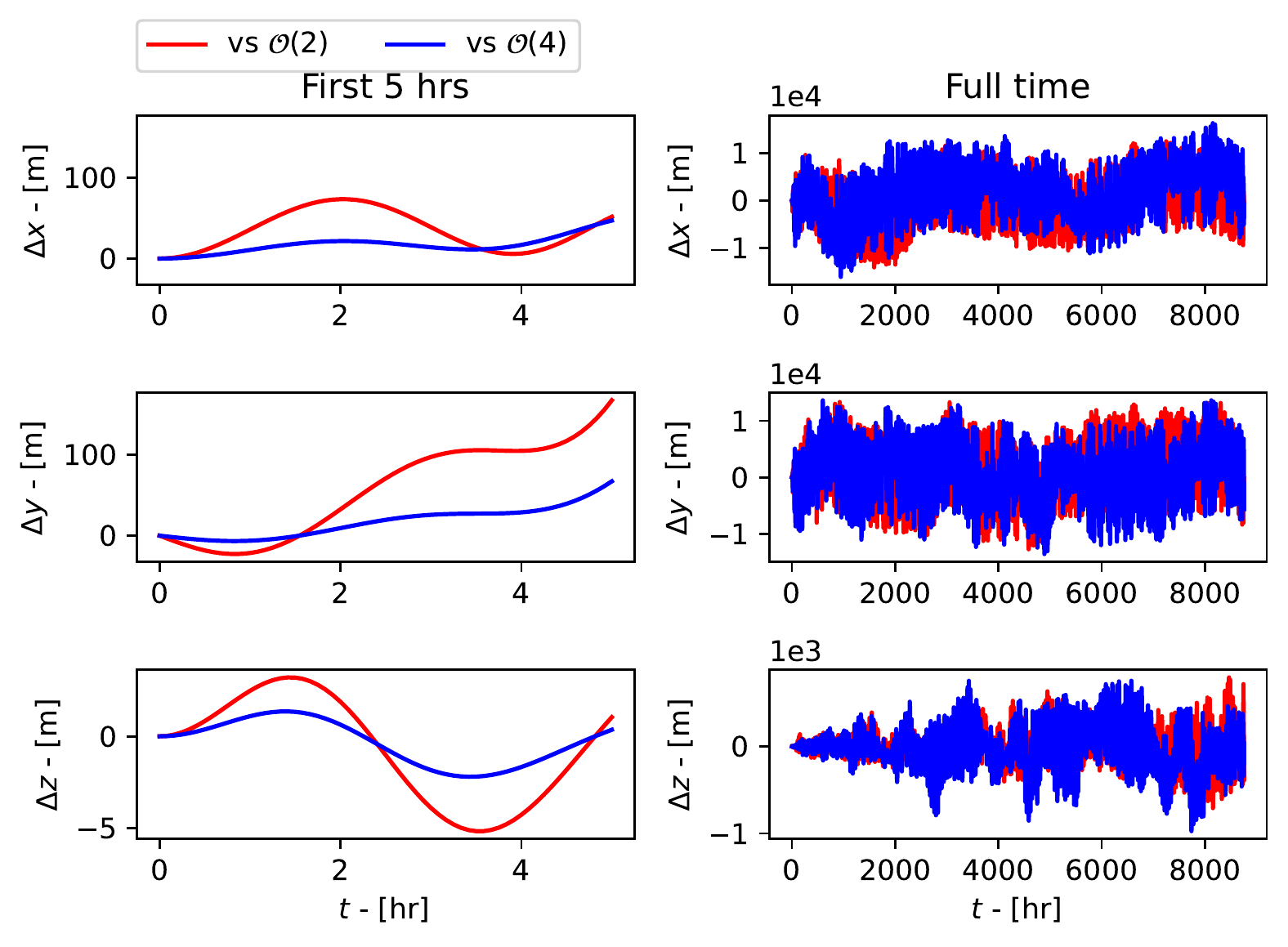}
\caption{For scenario 2: The difference in the $x$-, $y$- and $z$-coordinates of the position of the secondary relative to the primary. The red and blue lines correspond to the difference between the surface integration method and the order two and four approximations by \texttt{gubas}, respectively. 
The \textit{left column} shows the difference during the first five hours, while the \textit{right column} shows the difference over the whole simulation time span.}
\label{fig:Rotfission_position_diff}
\end{figure}

With these initial conditions, we compute again the difference in $x$-, $y$- and $z$-coordinates of the secondary as a function of integration time, and plot the result
in Fig.~\ref{fig:Rotfission_position_diff}, where the first five hours are plotted separately in the left-hand panels. After the first five hours, the coordinates already deviate by more than 50 m along the $x$ and $y$-directions. As time passes, the difference increases, and $\Delta x$ and $\Delta y$ can become larger than 10 km, and $\Delta z$ larger than 1 km. It also appears that the magnitude of the difference is the same regardless of whether the second order or the fourth order approximation is used. 

In order to check if the large differences could be caused by slightly different initial angular velocities, we started some simulations with the same initial angular velocity of $\omega_0 = 2.99 \cdot 10^{-4}$ rad/s (the value computed from the surface integration method) for all three models, but found that $\Delta x$-, $\Delta y$- and $\Delta z$ reach the same order of magnitude as that shown in Fig.~\ref{fig:Rotfission_position_diff}. Using $\omega_0 = 2.92 \cdot 10^{-4}$ rad/s as the initial angular velocity (the value from the second order potential) results in the bodies colliding for both the surface integration method and the fourth order approximation, already after the first time step.

The secondary orbits closer to the primary with the second order approximation, and the separation between the primary and the secondary is $\sim 4.5$ primary radii on average, and never exceeds eight primary radii. For the fourth order approximation and the surface integration method, the separation is on average $\sim 5.4$ and $\sim 5.5$ primary radii, respectively, and can reach up to ten primary radii. This is a consequence of the different initial conditions, as the second order potential yields a lower initial velocity compared to the other two methods. On the other hand, if the same initial conditions are used, the average separation is approximately $5.4$ primary radii for all three models. 

\begin{figure}
\centering
\includegraphics[width=\linewidth]{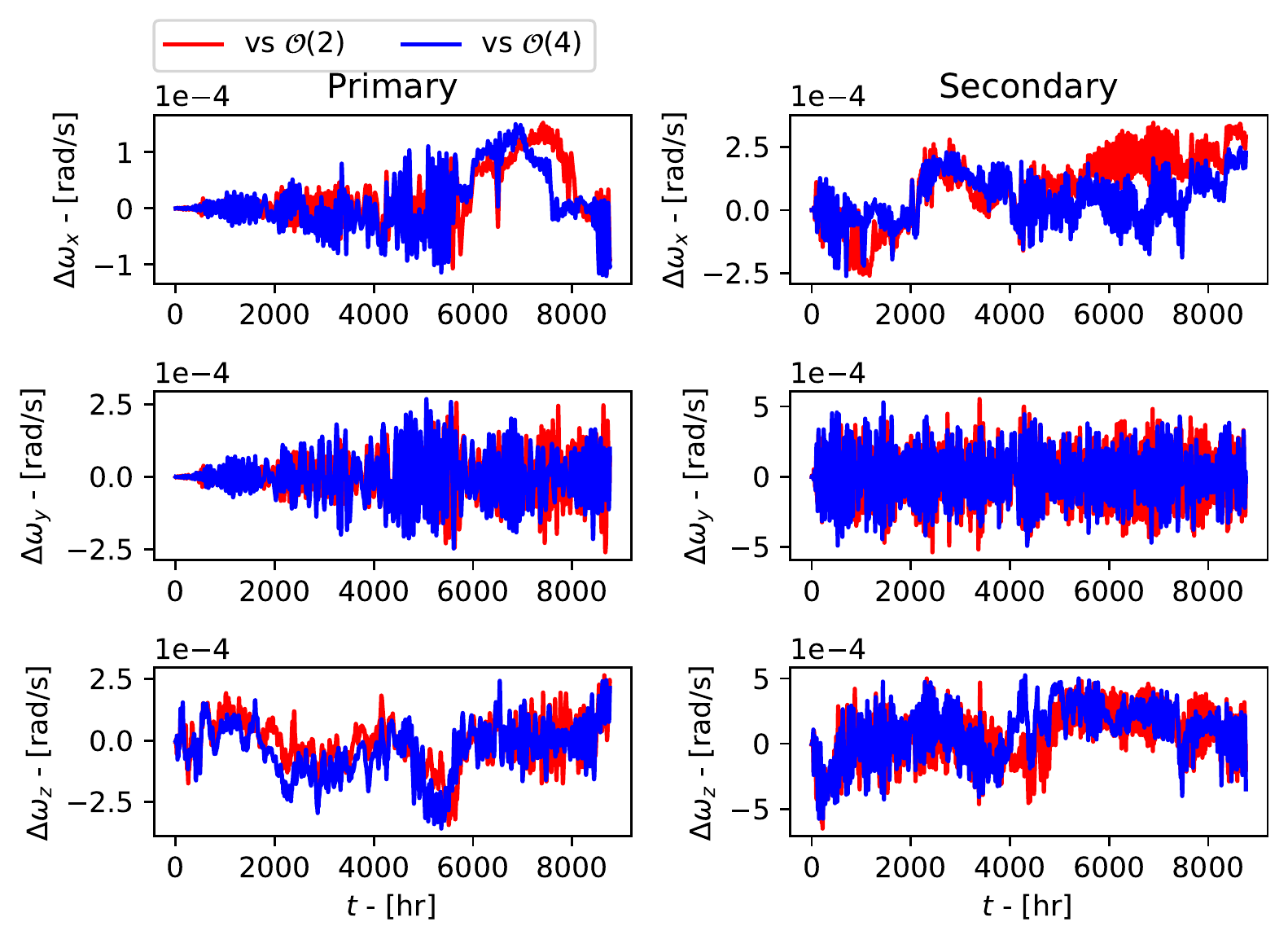}
\caption{The difference in angular velocity components for the primary (\textit{left}) and secondary (\textit{right}). The red and blue lines correspond to the difference between the surface integration method, and the order two and four approximations by \texttt{gubas}, respectively.}
\label{fig:Rotfission_angvel_diff}
\end{figure}

The differences in the angular velocity components are also larger compared to scenario 1, with a magnitude of $\Delta \omega$ of the order $10^{-4}$ rad/s for both bodies and for both approximations, shown in Fig.~\ref{fig:Rotfission_angvel_diff}. This difference also occurs when the same initial conditions are used.

\begin{figure}
    \centering
    \includegraphics[width=\linewidth]{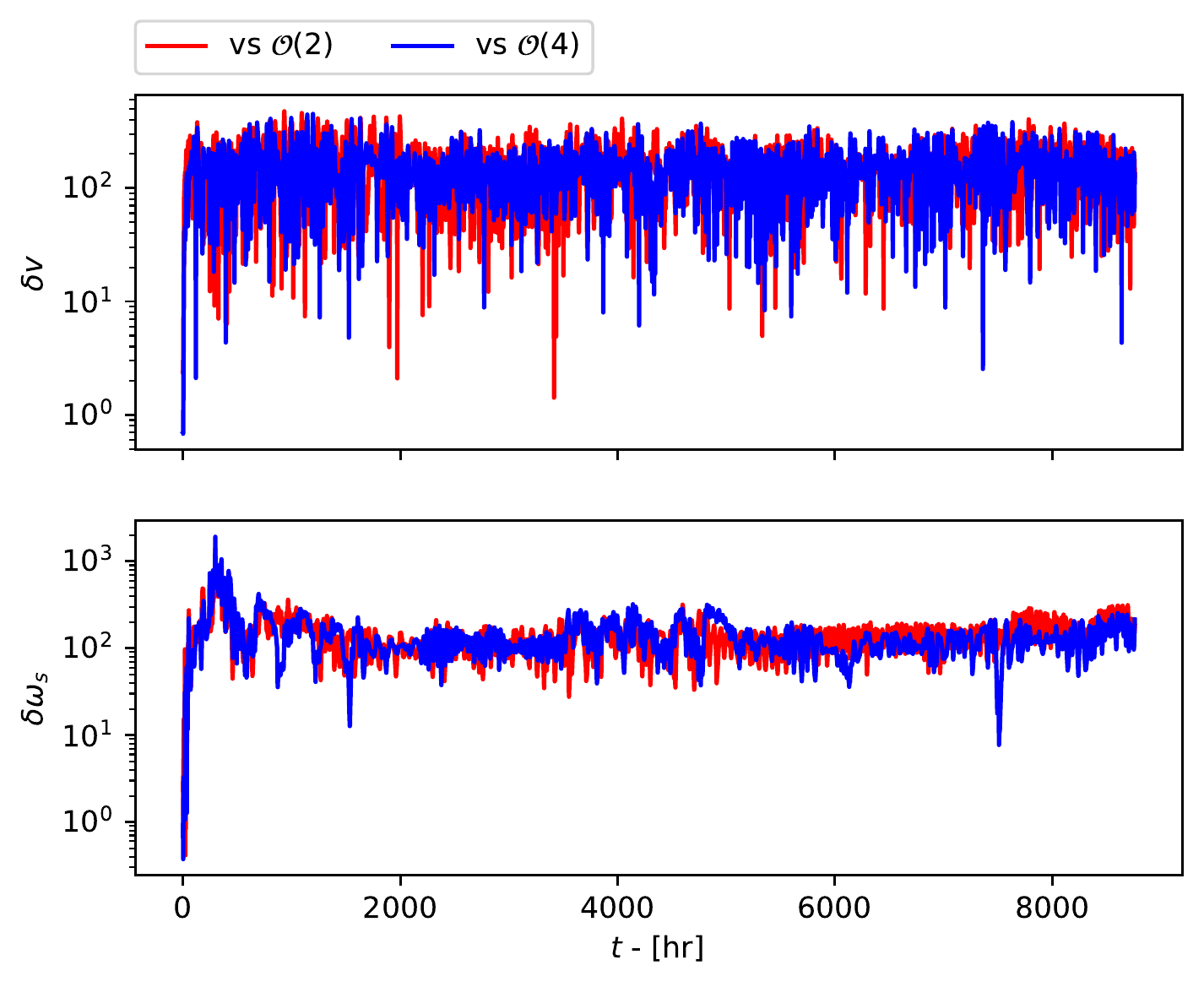}
    \caption{Same as Fig.~\ref{fig:Basic_demo_reldiffs}, but now for scenario 2.}
    \label{fig:Rotfission_reldiff_velocities}
\end{figure}
Contrary to the simulation in the previous subsection, where the bodies are further apart, both the second and the fourth order approximation produce equally erroneous velocities and angular velocities, as seen in Fig.~\ref{fig:Rotfission_reldiff_velocities}. The relative differences are greater than $130$\% on average, regardless of expansion order. Provided that the bodies are modeled as ellipsoids, using an exact method therefore becomes more important for the outcome of both the translational and rotational motion of the bodies if that they are initially very close.

The relatively large differences in the evolution of the binary between the surface integration scheme and the two other methods are likely due to the proximity of the bodies during the first few hours of the simulation. After about ten hours, the average separation between the bodies is sufficiently large that the difference in the mutual potential between the surface integration approach and the other two methods is relatively small, but as the bodies have evolved very differently up till then, they continue to evolve differently.  

\begin{figure}
\centering
\includegraphics[width=\linewidth]{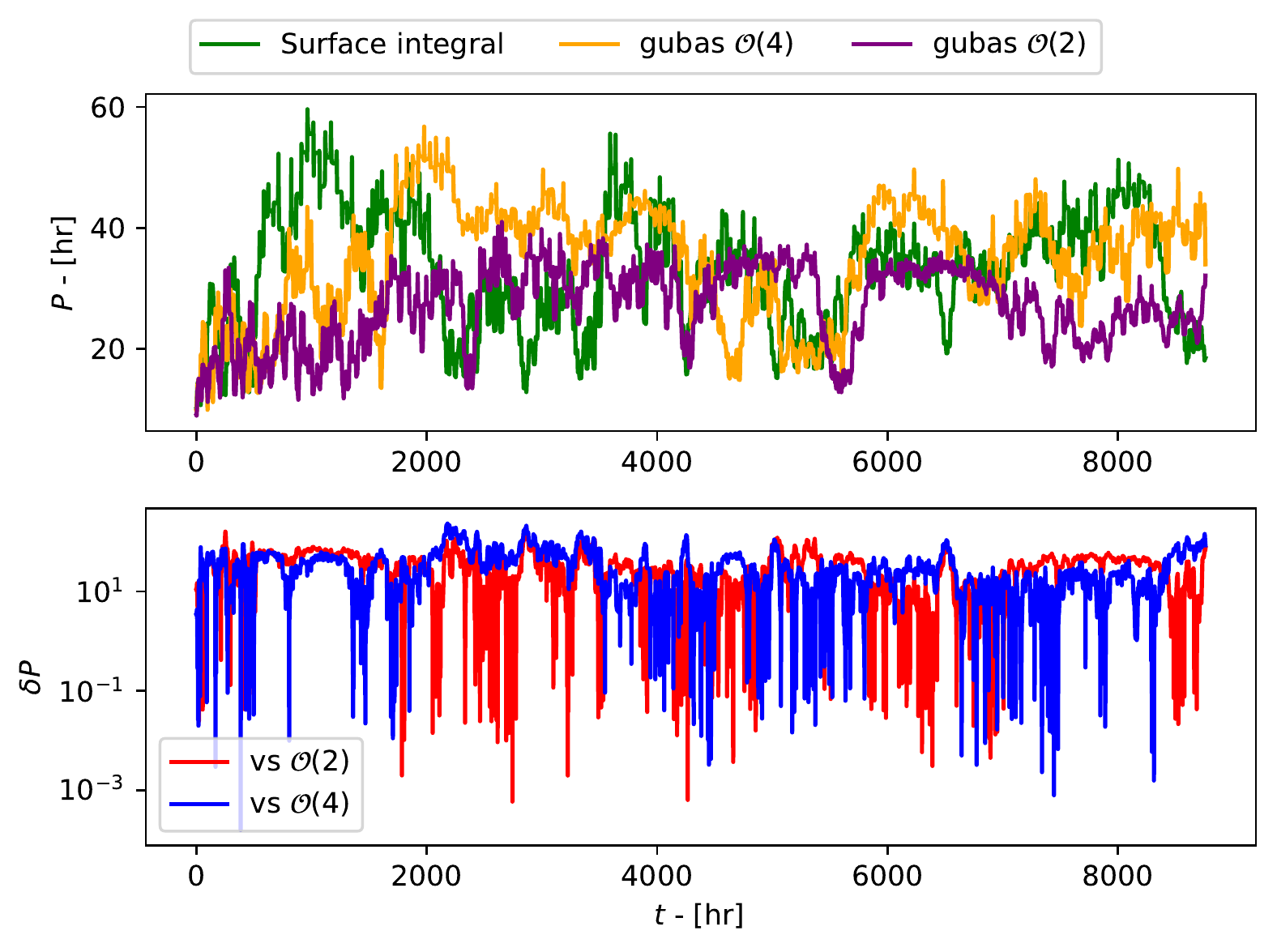}
\caption{For scenario 2: \textit{Top}: The orbital period of the secondary from the three models as functions of time. \textit{Bottom}: The relative difference (in percentage) in the orbital period between the expansion methods and the surface integration method.}
\label{fig:Rotfission_orbital_period}
\end{figure}

In Fig.~\ref{fig:Rotfission_orbital_period} we show how the
orbital period of the secondary changes over time with the three different methods. By using the second order potential, the orbital period is generally shorter compared to using either the fourth order potential or the surface integration method method, consistent with the secondary orbiting closer to the primary in the former case. The error in the orbital period can exceed $10$\% for both the second and fourth order potentials.

\citet{2017CeMDA.127..369H} find that higher order terms become important to determine the trajectories of post-fissioned binaries if the bodies are more elongated. In order to check whether changing the shape of the primary to a more spherical shape has large effects on the outcome, we changed the semiaxes of the primary to $(a_p, b_p, c_p) = (1000, 900, 850)$ m (which also results in different semiaxes and positions of the secondary, and different values of $\omega_0$), the discrepancy of the positions and angular velocities, between the surface integration method and the output of \texttt{gubas}, are of the same order of magnitude as in Figs. \ref{fig:Rotfission_position_diff} and \ref{fig:Rotfission_angvel_diff}. Therefore, the use of a more accurate method is also important for the dynamics of newly fissioned contact binaries, even if the bodies have low elongations.

\subsection{Energy differences - formation of asteroid pairs and stable binaries}
After a contact binary has separated into two components by rotational fission, the secondary may end up in a stable orbit around the primary, or escape (re-impact with the primary is also possible). This depends on the total energy of the system,  if the total energy is negative, the system may become a stable binary, whereas if the total energy is positive, and there is no loss of energy, the components may undergo mutual escape. Therefore, if we assume that there is no exchange of energy with the surroundings and that the bodies are rigid, the initial total energy of the system determines whether the contact binary becomes an asteroid pair or a binary \citep{2010Natur.466.1085P, 2011Icar..214..161J, 2016MNRAS.461.3982B, 2022alexpaper}.

In the rotational fission model, the initial energy of the contact binary depends on the mutual gravitational potential, and hence affects predictions of which systems may form binaries and which may form asteroid pairs. In order to address the influence that the choice of approximation order has on the ability to form stable binaries we compute the total energy for a number of different configurations with the three methods.

We assume a contact binary where the two components have equal shapes (as defined by their axis ratios), and vary the mass ratio from 0.01 to 0.3 as this is the region around zero total energy. For each assumed value of $q$, we choose several different orientation angles $\theta_s$ of the secondary, from  $0^\circ$ to $90^\circ$ (see Fig.~\ref{fig:RotFission_initial_illustration}). The initial conditions are the same as described in Sect. \ref{subsec:Rotfission_sim} \citep[see also][]{2022alexpaper}.

\begin{figure}
\centering
\includegraphics[width=\linewidth]{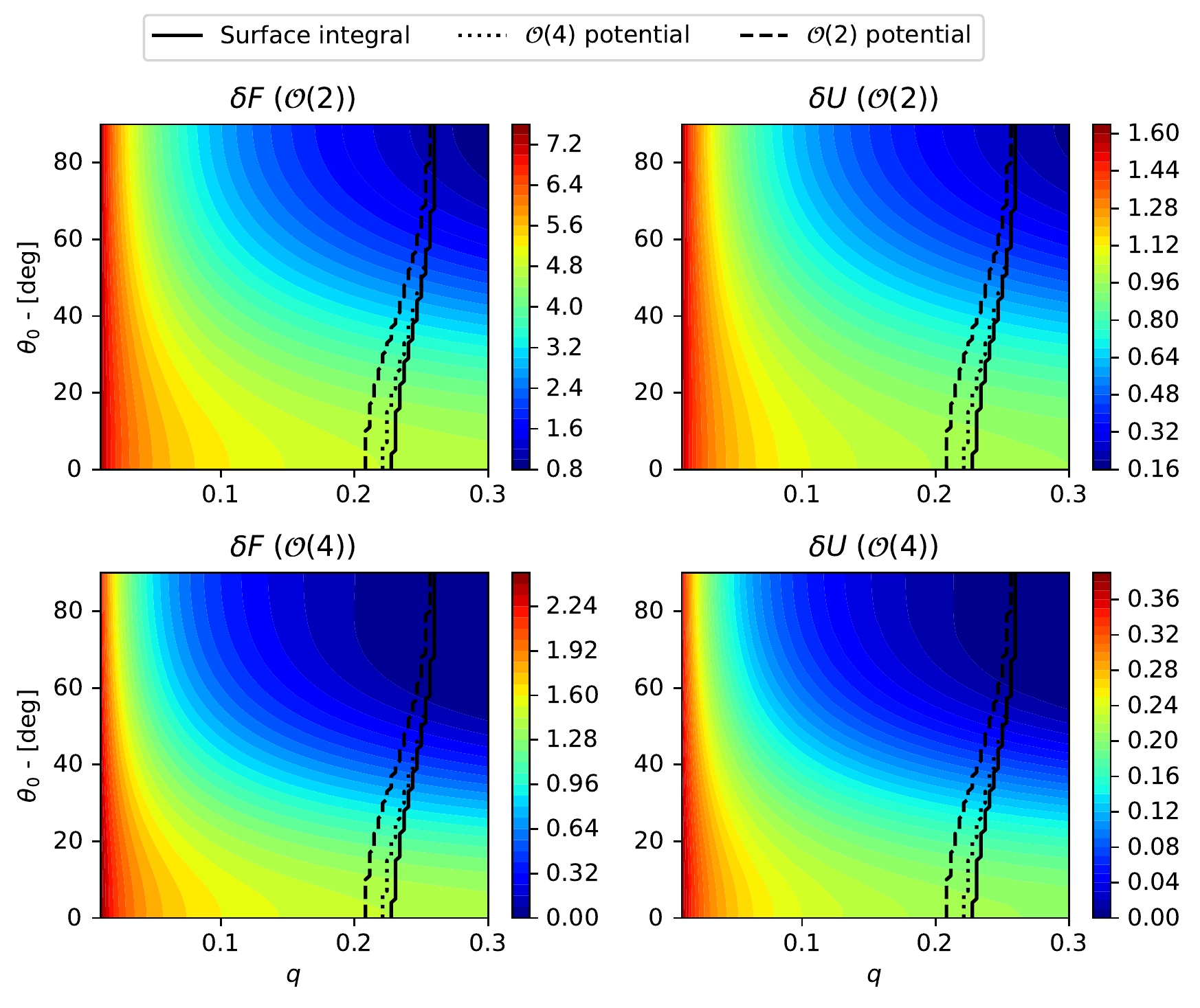}
\caption{Relative error (in per cent) in force and mutual potential energy in the \textit{left} and \textit{right} columns, respectively. The \textit{top} and \textit{bottom} rows correspond to the relative difference of the order two and order four potentials. The black lines indicate the separation between positive and negative total energies. Regions to the left and right of the respective lines correspond to positive and negative system energies.}
\label{fig:ForcePotential_rotfission_difference}
\end{figure}

The results are displayed in Fig.~\ref{fig:ForcePotential_rotfission_difference} in the form of a line in the $q$-$\theta_s$ plane marking the separation between positive and negative system energies. As the mass ratio increases, the total energy of the system starts to become negative but can remain positive if $\theta_s$ is large enough. The separation between positive and negative energies varies between the methods but generally ranges between $q \sim 0.21$ and $q\sim 0.26$. The fourth order approximation is seen to produce results that are fairly close to that from the surface integration method.

The difference between the methods is larger for lower values of $\theta_s$, but is still quite small, i.e. for a given $\theta_s$ the zero energy line occurs over a span in $q$ that is always less than 0.02. As seen in the figure, the fourth order and the surface integral method is quite similar, whereas the second order potential yields a zero energy line more shifted toward lower mass ratios. For instance, if the mass ratio is $q\approx 0.23$, the second order method predicts that the contact binary becomes a pair if $\theta_s \gtrsim 40^{\circ}$. But with the fourth order method, the contact binary can still form a stable binary as long as $\theta_s$ does not exceed $\sim 20^{\circ}$. Simulations that use the surface integral method for determining the mutual potential will therefore predict formation of asteroid pairs at slightly higher mass ratios compared to methods that use expansions of the mutual potential. 

As $\theta_s$ approaches 90 degrees, the separation vector $\mathbf{r}$ becomes smaller in order to maintain the 1 cm surface-to-surface separation. As a consequence, the bounding spheres between the bodies will intersect more when $\theta_s$ increases, and we expect the mutual potential to differ more between the methods. However, as seen in Fig.~\ref{fig:ForcePotential_rotfission_difference}, this is not the case, since the gap between the separation lines becomes smaller. Upon further inspection, the discrepancy in both the force and mutual potential energy become smaller when the system approaches higher mass ratios ($q\approx0.3$) with $\theta_s=90^{\circ}$, illustrated by the colored contours in Fig.~\ref{fig:ForcePotential_rotfission_difference}. The largest differences are found at low mass ratios and when $\theta_s$ is small, where the errors in the force can reach 7\% when $q=0.01$ for the second order potential, while error reduces to roughly 2\% when the potential is truncated to order four.

\section{Computational efficiency }
\label{sec:CPU_comparison}
While the surface integration method is exact for bodies of ellipsoidal shapes, it is also more time-consuming to compute as multiple double integrals must be solved and transcendental functions need to be evaluated. In this section, we compare the CPU times required to compute the forces and torques. We also compare the CPU times of the full dynamical simulations to complete. The comparisons are performed using the same single-core computer.

We first investigate the efficiency in the force and torque calculations. The evaluation times are measured in the code segments where the forces and torques are calculated, which excludes the time required to initialize the program and to solve the equations of motion. The second column of Table \ref{tab:Simulation_time} shows the CPU time required to evaluate the forces and the torques of both bodies, averaged over 37182 different configurations. The second and fourth order potentials are approximately 82 times and 4 times faster than the surface integration method, respectively, while the potential truncated to order 8 is roughly 16 times slower than the surface integration scheme.

The third and fourth columns of Table \ref{tab:Simulation_time} shows the CPU times of the simulations described in Sect. \ref{sec:Sim_comparisons}. Here, the CPU times are measured from the moment the respective programs initiate until they terminate. This includes the time required to initialize the program, solve the equations of motion, and saving the results. The output is saved at each time step both for our method and with \texttt{gubas} in order to facilitate the comparison, although this can be changed for \texttt{gubas} to reduce the CPU time. The second order potential used by \texttt{gubas} is very efficient compared to the surface integration method. However, if the potential is truncated to order four, the CPU times for the simulations are comparable to the surface integration scheme. This is due to the differences in how the equations of motions are solved, and how they are optimized, for each software. Finally, for higher orders of the potential it seems that the surface integral method would be preferable, as an approximation order of eight with \texttt{gubas} takes approximately 33 hours compared to 1369 seconds with the surface integral method.

\begin{table*}
\centering
\begin{tabular}{|c|c|c|c|}
\hline
Method & Force computations & Scenario 1  & Scenario 2\\
& Average CPU time [seconds] & CPU time [seconds] & CPU time [seconds] \\
\hline
Surface integral method & 0.055 & 1369 & 6435\\
\texttt{gubas} ($\mathcal{O}(2)$ expansion) & 6.723 $\cdot 10^{-4}$ & 96 & 375\\
\texttt{gubas} ($\mathcal{O}(4)$ expansion) & 0.014 & 1555 & 6120\\
\texttt{gubas} ($\mathcal{O}(8)$ expansion) & 0.877 & 119474 & \\
\hline
\end{tabular}
\caption{Comparison of CPU times for the methods. The second column shows the average CPU time required to compute the forces and torques. Scenario 1 and 2 correspond to the simulations presented in Sect. \ref{sec:Sim_comparisons}, where the first one is for a binary with larger separation, and the second one is for a newly fissioned contact binary.}
\label{tab:Simulation_time}
\end{table*}

\section{Summary and discussion}
\label{sec:Discussion}
By utilizing the surface integral method that we have developed and described in some recent publications \citep{2021CeMDA.133...27W, 2021CeMDA.133...35H, 2022alexpaper}, we are able to accurately describe, without approximations, the gravitational interaction between two triaxial ellipsoids. This makes us able to address errors in force and torque calculations between two ellipsoids using methods based on series approximations, such as the inertia integral method, where the mutual gravitational potential is truncated at a certain order. A publicly available implementation of this is \texttt{gubas} \citep{DAVIS2020113439}, which we use in this work to compute interactions between two ellipsoids based on potentials truncated at second and fourth order. Previous work have compared approximative methods to each other for ellipsoidal \citep{2017CeMDA.127..369H} and polyhedral shapes \citep{2020Icar..34913849A}. In this manuscript, we have compared the surface integration method with a method that expands the mutual potential truncated up to order four. 

For a typical binary asteroid, where the secondary orbits the primary in the equatorial plane, both the second and fourth order potentials give relatively good approximations to the more realistic value of the force, the error being insignificant at distances of 3--5 primary radii, and less than one per cent even very close to the primary. 

For a typical binary asteroid, where the secondary orbits the primary in the equatorial plane, both the second and fourth order potentials give similar values of the force compared to the surface integration method. The errors become insignificant at distances of 3--5 primary radii, and less than one percent even when the secondary is close to the primary.

For the torque, however, the errors become more significant, especially if the bodies are displaced such that the separation vector between the mass centers, $\mathbf{r}$, is parallel with one of the principal axes of the body for which the torque is being calculated. In this case the second order approximation fails by 100\%. This is due to a mathematical limitation inherent in the second order approximation \citep{kane1983spacecraft, 2012JCoPh.231.7237P}. Fourth order potentials can correct somewhat for this, but generally if the other body lies in the neighbourhood of one of the principal axis of the body for which the torque is evaluated, the errors in the torque are notably larger than elsewhere. Consequently, approximative methods affect rotational motion more than translational motion, and using a more accurate method therefore becomes more important to correctly describe the rotational motion. The percentage errors in the torques are approximately an order of magnitude larger than the errors in the force. However, as long as the separation between the two components of the binary is sufficiently large (a few primary radii), simulations using the surface integration method and the expansion approaches show negligible differences in the torques. 

The most notable differences and largest errors occur in situations where the two bodies are close with their centroids not in the same plane. These configurations are particularly relevant for contact binaries that separate due to rotational fission when a certain spin limit is reached. The two bounding spheres of the bodies intersect, and the series approximation of the mutual potential described by Eq.~\eqref{eq:Hou_mutual_potential} no longer converges \citep{2008CeMDA.100..319T}. The surface integration approach, on the other hand, is still valid, and we find that for a secondary placed close to the surface of the primary (insignificant surface-to-surface distance of 1 cm), errors are largest when the secondary is placed closed to the pole of the primary. The errors from the second order approximation are generally larger than from the fourth order, but if the primary is elongated enough in this configuration, the errors from the fourth order approximation may dominate. This is because the forces obtained by inertia integrals scale as $(a_p/r)^N$ for an order $N$ expansion. When $r < a_p$ (as is the case for overlapping bounding spheres), the contribution from higher order gravity terms may inflate the calculated force, and lead to large errors.

Using a more accurate method to determine the forces and torques in the initial stages becomes more important if the bodies are initially close and the bounding spheres overlap. In these cases, the difference in the computed forces between the methods result in significantly different angular and translational velocities in the initial stages of the simulation. This leads to, for a binary with $q=0.3$, deviations in the position of the secondary relative to the primary of more than 10 km, while the angular velocity components differ by $\sim 10^{-4}$ rad/s. The latter corresponds to relative differences in the angular velocity (and rotation period) that exceed 100\%. We also find that these initial differences can lead to differences in the orbital period of the secondary of more than $10$\%. These discrepancies in the simulations are also seen even when the initial conditions are equal, which indicates that the use of a more accurate method to determine the mutual potential is particularly important the first hours of a post-fissioned asteroid system, in which the bodies are still relatively close.

With the surface integration method we calculate higher system energies for contact binary systems, compared to second and fourth order methods. Assuming that there is no loss of energy our calculations therefore predict formation of asteroid pairs (total energy positive) at slightly higher mass ratios. These calculations were done for a few configurations where the secondary lies in the equatorial plane of the primary, but the separation between positive and negative system energies also changes with the shape of the bodies and their densities \citep{2022alexpaper}. Furthermore, the outcome of the system, e.g.\ whether the two components collide or how long the secondary remains in orbit before escaping, may also be affected by how the mutual potential is computed. A future study comparing these outcomes may further demonstrate the importance of using a more accurate method to determine the mutual potential in order to study dynamics of post-fissioned asteroid systems.

The mass ratio of the system does not significantly alter the differences in the computed forces or torques between the methods, provided that the bodies are sufficiently far apart. However, if the bodies are close, the use of a more realistic method becomes more important to the mutual potential for systems with lower mass ratios. For a post-fissioned asteroid system, the error in the force, from a second order potential, can reach $\sim 7$\% when $q = 0.01$, and reduces to $\sim 4$\% when $q = 0.30$. When the potential is truncated to order four, the errors in the force are reduced to $\sim 2$\% and $\sim 1$\% for mass ratios $0.01$ and $0.30$ respectively.  

We benchmark the methods by comparing the CPU times required to compute the forces and torques, and to complete the long-term simulations. Due to the nature of double integrals, the forces computed by the surface integration scheme is slower than that of \texttt{gubas}. However, for the full simulations, the time required for the simulations to finish from the surface integration method is comparable to the ones from \texttt{gubas} when an order four potential is used.

In this manuscript, we have only considered bodies of ellipsoidal shapes. However, modeling an asteroid as a polyhedron provides a more realistic representation of its shape. It may be of interest to compare the surface integration method with other expansion methods, applied to polyhedral shapes in the future.

\section*{Acknowledgements}
The authors would like to thank Sverre Lunøe-Nielsen for helpful discussions on numerical issues. We would also like to thank the anonymous referee for their feedback that improved the manuscript.


\appendix
\section{Open-source software}
The software that uses makes use the surface integration method, as outlined by \citet{Conway_2016} and \citet{2021CeMDA.133...35H}, to solve the F2BP is available in the following GitHub repository \url{https://github.com/alexhosians/SIANS}. 

\bibliography{references} 

\end{document}